\documentclass[twocolumn, twocolappendix, tighten, times]{aastex63}

\usepackage{epstopdf}
\usepackage{amsmath}
\usepackage{enumitem}

\newcommand{\HD}{HD 107146}
\newcommand{\HDN}{HD 92945}
\newcommand{\HDO}{HD 15115}
\newcommand{\HDnew}{HD 206893}
\newcommand{\Alowerep}{\texttt{A-Loep}}
\newcommand{\Ahigherep}{\texttt{A-Hiep}}

\def\inn{_{\rm in}}
\def\out{_{\rm out}}
\def\res{_{\rm res}}
 
\shorttitle{Sculpting gaps in massive debris disks}
\shortauthors{A. A. Sefilian, R. R. Rafikov, and M. C. Wyatt}

\begin{document}

\title{Formation of Gaps in Self-Gravitating Debris Disks by Secular Resonance in a Single-planet System I: A Simplified Model}

\correspondingauthor{Antranik A. Sefilian}
    \email{aas79@cam.ac.uk, sefilian.antranik@gmail.com}

\author[0000-0003-4623-1165]{Antranik A. Sefilian}
    \affil{Department of Applied Mathematics and Theoretical Physics, University of Cambridge, Wilberforce Road, Cambridge CB3 0WA, UK}

\author[0000-0002-0012-1609]{Roman R. Rafikov}
    \altaffiliation{John N. Bahcall Fellow at the Institute for Advanced Study}
    \affil{Department of Applied Mathematics and Theoretical Physics, University of Cambridge, Wilberforce Road, Cambridge CB3 0WA, UK}
    \affiliation{Institute for Advanced Study, Einstein Drive, Princeton, NJ 08540, USA}
 
\author[0000-0001-9064-5598]{Mark C. Wyatt}
    \affil{Institute of Astronomy, University of Cambridge, Madingley Road, Cambridge CB3 0HA, UK}

\begin{abstract}
\noindent 
Spatially resolved images of debris disks frequently reveal complex morphologies such as gaps, spirals, and warps. Most existing models for explaining such morphologies focus on the role of massive perturbers (i.e. planets, stellar companions), ignoring the gravitational effects of the disk itself. Here we investigate the secular interaction between an eccentric planet and a massive, external debris disk using a simple analytical model. Our framework accounts for both the gravitational coupling between the disk and the planet, as well as the disk self-gravity -- with the limitation that it ignores the non-axisymmetric component of the disk (self-)gravity. We find generally that even when the disk is less massive than the planet, the system may feature secular resonances within the disk (contrary to what may be naively expected), where planetesimal eccentricities get significantly excited. Given this outcome we propose that double-ringed debris disks, such as those around \HD~and \HDN, could be the result of secular resonances with a yet-undetected planet interior to the disk. We characterize the dependence of the properties of the secular resonances (i.e. locations, timescales, and widths) on the planet and disk parameters, finding that the mechanism is robust provided the disk is massive enough. As an example, we apply our results to \HD~and find that this mechanism readily produces $\sim 20$ au wide non-axisymmetric gaps. Our results may be used to set constraints on the total mass of double-ringed debris disks. We demonstrate this for \HDnew, for which we infer a disk mass of $\approx 170 M_{\earth}$ by considering perturbations from the known brown dwarf companion.
\end{abstract}


\section{Introduction}   
\label{sec:intro}

Debris disks are ubiquitous around main sequence stars, with current detection rates of $\sim 20$\% in the Solar neighbourhood \citep{montesinos2016, sibthorpe2018}. They are optically thin, almost devoid of gas, and are believed to be composed of objects ranging from micron-sized dust grains up to kilometre-sized planetesimals. Since the dust grains are short-lived compared to the stellar age \citep[e.g.][]{dd03}, their sustained presence requires a massive reservoir of large planetesimals continually supplying fresh dust via mutual collisions \citep{backman93}. Observed disks typically contain $0.01-1M_{\earth}$ in mm/cm-sized grains \citep{wyattscuba, holland17} which, when extrapolated, yields masses of $\sim 1-100M_{\earth}$ for the parent planetesimal population \citep[e.g.][]{wyattdent2002, greaves2005, krivovwyatt20}.  The spatial distribution of these planetesimals is probed indirectly with observations at millimetre wavelengths, e.g. by ALMA. At such wavelengths, observations trace the distribution of mm-sized dust which are largely insensitive to radiation forces, thus serving as proxy for the distribution of parent planetesimals.

Recent high-resolution observations of debris disks by ALMA and direct imaging have revealed a rich variety of radial and azimuthal structures: e.g., gaps or double-ringed structures, warps, spirals, and eccentric rings \citep[e.g.][]{hughes2018review, wyatt18review, wyatt19review}. Analogous to the studies of the asteroid and Kuiper belts, investigating the structure of debris disks can provide unique insight into the architecture and evolution of exoplanetary systems. For instance, the presence of a giant planet around $\beta$ Pictoris, dubbed as $\beta$-Pic b, was predicted based on the warp in the debris disk \citep{betapic97}, and such a planet was later discovered by direct imaging \citep{lagrange2010}. As such, modelling of disk morphology is often focused on investigating the dynamical imprints of (invoked) massive perturbers, e.g. planets \citep[e.g.][]{wyattetal99, wyattSPIRAL05, leechiang2016} or stellar companions \citep[e.g.][]{nesvold17}.

However, studies of planet-debris disk interactions usually ignore the gravitational effects of the disc itself. That is, debris disks are treated as a collection of massless particles subject only to the gravity of the star and (putative) planets. Nonetheless, this assumption may not always be justified, especially in view of observations suggesting that debris disks could contain tens of Earth masses in large planetesimals \citep{wyattdent2002, greaves2005, krivovwyatt20}. In this regard, \citet{jalali12} have argued that many of observed debris disk features could be ascribed to the slow ($m=1, ~ 2$) modes which, if and when excited (e.g. by stellar flybys), could be supported by the disk gravity alone. Despite this fact, gravitational effects of debris disks have not yet been widely appreciated in the literature.

In this paper (the first in a series) we investigate the interaction between an eccentric planet and an external, massive debris disk. The primary aim of this work is to present a novel pathway to sculpting gaps, i.e. depleted regions, in broad debris discs.

\subsection{Existing mechanisms and this work} 
\label{subsec:overview}

To date, four debris disks are known to exhibit double-belt structures that are separated by depleted gaps in their dust distribution as traced by ALMA: \HD~\citep{ricci2015, marino2018}, \HDN~\citep{marino2019}, \HDO~\citep{macgregor2019}, and \HDnew~\citep{marino2020}. These systems (except \HDnew) have no known companions or planets to date, and the disks are gas-poor. In this work we focus on the nearly face-on disk of \HD, a nearby $\sim$80-200 Myr old G2V star \citep{williams2004}. This disk, extending from $\sim$30 au to $\sim$150 au, features a circular $\sim$40 au wide gap centred at around $70-80$ au in which the continuum emission drops by $\sim 50\%$ \citep{ricci2015, marino2018}.

Various mechanisms have been explored for explaining the origin of gaps in debris disks. In analogy with the asteroid and Kuiper belts, the most popular scenario involves the presence of single or multiple planets orbiting within the depleted region, which are either stationary or migrating \citep[e.g.][]{christian2016, shannon2016, zheng2017, kaitlin2018}. 
For instance, it has been suggested that multiple stationary planets or a single but migrating planet of few tens of Earth masses on a near-circular orbit at $\sim70-80$ au could reproduce \HD's gap \citep[e.g. see][]{ricci2015, marino2018}.

Other scenarios involving planets interior to the disk, rather than embedded within, have also been considered. For instance, \citet{maryam} showed that a low-eccentricity planet can carve a gap at its external 2:1 mean motion resonance. On the other hand, \citet{pearcewyatt15} demonstrated that \HD-like disks could be produced as a result of secular interactions and scattering events between a massive ($\sim 10 - 100 M_{\earth}$) planetesimal disk and an initially high-eccentricity ($\sim 0.5$) planet of comparable mass to the disk. In the course of evolution, the planetary orbit is then circularized due to scattering events. However, \citet{pearcewyatt15} consider only the back reaction of the disk on the planet (and vice versa) in their simulations, neglecting the disk self-gravity.

Finally, \citet{yelverton2018} considered a scenario whereby two coplanar planets carve a gap through their secular resonances within an external debris disk, which was assumed to be \textit{massless}. In their model, the secular resonances occur at sites where the precession rates of the planets (i.e. system's eigenfrequencies) match that of the planetesimals in the disk (due to planetary perturbations). They find that at and around \textit{one} of the two resonant sites planetesimal eccentricities are excited, triggering a depletion in the disk surface density of the kind seen in \HD.

The model proposed by \citet{yelverton2018} requires (at least) two planets to ensure that their orbits are precessing due to planet-planet interactions, a condition necessary for establishing secular resonances. However, another mechanism which may drive planetary precession is the secular perturbation due to the disk, which was ignored by \citet{yelverton2018}. This motivates our investigation into whether gaps could be carved in \textit{self-gravitating} debris disks via secular resonances when perturbed by single rather than multiple inner planets. A related scenario was studied by \citet{zheng2017} which showed that a single planet embedded \textit{within} a decaying gaseous disk (i.e. transitional disk) could carve a wide gap around its orbit via sweeping secular resonances assisted by the waning disk gravity.

In this paper we propose that double-ringed structures -- akin to that of \HD~ -- could be explained as the aftermath of secular resonances in systems hosting a \textit{single} eccentric planet and an external self-gravitating debris disk. The mechanism we invoke here is different from those of \citet{pearcewyatt15} and \citet{yelverton2018}. It is realized through a secular resonance between the apsidal precession rate of planetesimals due to \textit{both} the disk and planet, and that of the planet due to \textit{disk gravity} \citep[c.f.][]{yelverton2018}. Additionally, our mechanism does not require scattering events between the planet and disk particles \citep[c.f.][]{pearcewyatt15}. As we show below, this mechanism is robust over a wide range of parameters; particularly when the disk is less massive than the planet.

Our work is organized as follows. In Section \ref{sec:theory} we describe our model system and present the equations governing planetesimal dynamics. In Section \ref{sec:resonances} we characterize the features of the secular resonances over a wide range of parameter space. In Section \ref{sec:HD_constraints} we apply these considerations to \HD, and identify the planet-disk parameters which could reproduce the observed gap. In Section \ref{sec:results_new}, using some of these parameters, we investigate the evolution of disk-planet systems and present our main results. We discuss our results along with their implications in Section \ref{sec:discussion}, where we also consider the application of our results to other systems (\HDN~and \HDnew). In Section \ref{sec:limitations_future} we critically assess the limitations of our model, discuss the implications of relaxing some of them, and propose future work.  Our findings are summarized in Section \ref{sec:summary}.

\section{Analytical Model} 
\label{sec:theory}

We describe a simple model to analyze the long-term dynamical evolution of planetesimals embedded within a massive debris disk in a single-planet system. In our notation, a planetesimal orbit is characterized by its semimajor axis $a$, eccentricity $e$, and longitude of pericenter $\varpi$. Orbital elements subscripted with `$p$' and `$d$' refer to the planet and the disk, respectively.

\subsection{Model system} 
\label{subsec:model_system}

Our model system consists of a broad debris disk of mass $M_d$ orbiting the host star $M_c$ exterior to, and co-planar with, a planet of mass $m_p$ ($M_d, m_p \ll M_c$).  We assume the planet is initially on a low eccentricity orbit ($e_p \leq 0.1$) and that it does not intersect the disk along its orbit. We consider the debris disk to be razor-thin and initially axisymmetric. The disk surface density is characterized with a truncated power-law profile given by
\begin{equation}
    \Sigma_d(a) = \Sigma_0 \left( \frac{a\out}{a} \right)^{p} 
    \label{eq:Sigma_d}
\end{equation}
for $a\inn \leq a \leq a\out$, and $\Sigma_d(a) = 0$ elsewhere. Here, $a\inn$ and $a\out$ are the semimajor axes of the inner and outer disk edges, respectively. Defining $\delta \equiv a\out/a\inn > 1$, the total mass $M_d$ of such a disk can be written as 
\begin{equation}
    M_d =  \frac{2\pi}{2-p} \Sigma_0 a_{\rm out}^2 \left( 1- \delta^{p-2} \right) ,
    \label{eq:disk_mass}
\end{equation}
which allows us to express $\Sigma_d$ in terms of $M_d$. This setup is very similar to that explored in \citet{RRptype} and \citet{silsbeekepler} in the context of planetesimal dynamics in circumbinary disks.

In this work, unless otherwise stated, we adopt a fiducial disk model with $p=1$, $a\inn = 30$ au and $a\out = 150$ au (i.e. $\delta = 5$). This choice of $p$ corresponds to a disk with constant amount of mass per unit semimajor axis.

\subsection{Secular gravitational effects}
\label{sec:dist_function}

We are primarily interested in the long-term dynamics of large ($\sim$km-sized) planetesimals. Since the latter are effectively insensitive to radiative non-gravitational forces, we focus purely on gravitational effects accounting for perturbations due to \textit{both} (1) the debris disk and (2) the planet. For simplicity, the non-axisymmetric component of the disk gravity is ignored in this work, although, as we will see later, the disk naturally develops non-axisymmetry (a discussion of the implications of this omission is provided in \S \ref{subsubsec:non_axi_effects}). We perform calculations within the framework of secular (orbit-averaged) perturbation theory to second order in eccentricities \citep{mur99}.

\subsubsection{Effects of the disk and planet on planetesimals}

The secular dynamics of planetesimals is described by the disturbing function $R$ which consists of contributions due to the planet $R_p$ and due to the disk $R_d$. An analytic expression for the disturbing function $R_d$ due to an axisymmetric disk with surface density of the form (\ref{eq:Sigma_d}) has been previously derived in \citet{sil15} \citep[see also][]{hep80,ward81, ST19}. Combining $R_d$ with the contribution $R_p$ due to the planet \citep[e.g.][equation 7.7]{mur99}, the total disturbing function $R = R_d + R_p$ to second order in eccentricities reads as: 
\begin{eqnarray}
    R &=&  n a^2 \left[ \frac{1}{2} A e^2 
    + B_p e \cos\left( \varpi - \varpi_p   \right) \right] ,
    \label{eq:RdRp}
\end{eqnarray}
where $n = \sqrt{G M_c / a^3}$ is the planetesimal mean motion and the meaning of different constants is explained below.

In Equation (\ref{eq:RdRp}), $A = A_d + A_p$ is the precession rate of the free eccentricity vector of a planetesimal. It has contributions from both the gravity of the disk $(A_d)$ and the planet ($A_p$). The contribution of the planet is \citep{mur99}
\begin{eqnarray}
    A_p &=& \frac{1}{4} n \frac{m_p}{M_c} \frac{a_p}{a}b_{3/2}^{(1)}(a_p/a) , 
    \label{eq:Aplanet}
    \\
    &\approx&  35.5 \times 10^{-2} ~ \mathrm{Myr}^{-1} \frac{m_p}{0.6 M_{J}} a_{p, 20}^2 ~ a_{70}^{-7/2} M_{c,1.09}^{-1/2}, 
    \nonumber
\end{eqnarray}
where $a_{p, 20} \equiv a_p/(20~\rm{au})$, $a_{70} \equiv a/(70~\rm{au})$, $M_{c,1.09} \equiv M_c/(1.09 M_{\odot})$, $b_s^{(m)}(\alpha)$ is the Laplace coefficient defined by 
\begin{equation}
    b_{s}^{(m)}(\alpha) = \frac{2}{\pi} \int\limits_{0}^{\pi} \frac{\cos(m\theta)d\theta}{(1+\alpha^2-2\alpha\cos\theta)^s},
\end{equation}
and the numerical estimate in Eq. (\ref{eq:Aplanet}) assumes $a_p/a \ll 1$ so that $b_{3/2}^{(1)}(\alpha) \approx 3\alpha$. The contribution of the disk to the free precession is \citep{sil15}
\begin{eqnarray}
    A_d  &=& 2\pi \frac{G \Sigma_d(a)  }{  n a   } \psi_1 =  (2-p)  n \frac{M_d}{M_c}  \bigg( \frac{a}{ a_{\rm out}  } \bigg)^{2-p} \frac{\psi_1}{ 1 - \delta^{p-2}  } 
    \label{eq:Adisc}
    \\
    &\approx &   - 14.4 \times 10^{-2} ~ \mathrm{Myr}^{-1}  \frac{M_d}{20 M_{\earth}} a_{70}^{-1/2} \frac{ M_{c,1.09}^{-1/2} }{a_{\rm{out}, 150}}
     \frac{\psi_1}{-0.5} 
    ,
    \nonumber  
\end{eqnarray}
where $a_{\rm{out}, 150} \equiv a\out/(150~\rm{au})$, and the numerical estimate is for $p=1$ and $\delta \gg 1$ such that $\psi_1 \approx -0.5$.

In general, the coefficient $\psi_1$ in Eq. (\ref{eq:Adisc}) depends on the power-law index $p$ as well as the planetesimal semimajor axis with respect to the disk edges \citep[][equation A33]{sil15}. As the sharp edges of the disk are approached, $\psi_1$ formally diverges. However, when the planetesimal is well separated from the edges (i.e. $a\inn \ll a \ll a\out$), $\psi_1$ is effectively a constant of order unity (depending on $p$) which can be well approximated by equation (A37) in \citet{sil15}. It is very important to note that the disk and the planet drive planetesimal precession in opposite directions, $A_p > 0$ and $A_d < 0$, with $A_p(a)$ falling off more rapidly with $a$ than $|A_d(a)|$.

The term $B_p$ in Eq. (\ref{eq:RdRp}) represents the excitation of planetesimal eccentricity due to the non-axisymmetric component of the planetary potential. It is given by \citep{mur99}
\begin{equation}
    B_p = -\frac{1}{4} n \frac{m_p}{M_c} \frac{a_p}{a} b_{3/2}^{(2)}(a_p/a) e_p .
    \label{eq:Bplanet}
\end{equation}
Note that the analogous term due to the disk is absent in Eq. (\ref{eq:RdRp}), since we have neglected  the non-axisymmetric component of the disk self-gravity.

\subsubsection{Effect of the disk on planet}

Next we consider the effect of the disk on the planet. Since the disk is taken to be axisymmetric, it simply causes the planetary apsidal angle to advance linearly in time such that $\varpi_p(t) = A_{d,p} t + \varpi_{p}(0)$, i.e. $\dot{\varpi}_p = A_{d,p}$, without exchanging its angular momentum with the planet. In this work, without loss of generality, we set $\varpi_p(0) = 0$. In Appendix \ref{app:effect_of_disk} we show that the planetary precession rate $A_{d,p}$ due to the disk with surface density (\ref{eq:Sigma_d}) is given by \citep[see also,][]{petrovich19}:
\begin{eqnarray}
    A_{d,p} &=& 
    \frac{3}{4} n_p \frac{2-p}{p+1}  \frac{M_d}{M_c}  
    \left(\frac{a_p}{a_{\rm out}} \right)^3
    \frac{\delta^{p+1} -1  }{ 1- \delta^{p-2}}  
    \phi_1^c 
    \label{eq:Adp}
    \\
    &\approx&  19.4 \times 10^{-2} ~ \mathrm{Myr}^{-1}  \frac{M_d}{20 M_{\earth}} 
    \frac{ a_{p,20}^{3/2} }{ a_{\rm{out}, 150} ~ a_{\rm{in}, 30}^2 }
    M_{c,1.09}^{-1/2} , 
    \nonumber 
\end{eqnarray}
where $n_p = \sqrt{G M_c / a_p^3}$ is the planetary mean motion, $a_{\rm{in}, 30} \equiv a\inn/(30~\rm{au})$, and the numerical estimate is for $p=1$ and $a_p = 20$ au such that $\phi_1^c \approx 1.8$. Here $\phi_1^c = \phi_1^c(a_p/a\inn, p, \delta)$ is a factor of order unity accounting for contributions of the disk annuli close to the planet (Eq. \ref{eq:phi1c_appendix}). Its behavior as a function of $a_p/a\inn$ and for various disk models (i.e. $p$, $\delta$) is shown in Fig. \ref{fig:phi_i_12_app}. For $a_p/a\inn \ll 1$, we have $\phi_1^c \approx 1$ regardless of $(p, ~ \delta)$.

\subsubsection{Combined planet-disk effects}

The fact that the planet is precessing renders the forcing term in $R$ (Eq. \ref{eq:RdRp}) time-dependent. This time dependence could be eliminated upon transferring to a frame precessing with the planetary orbit: i.e. by subtracting $\Phi A_{d,p}$ from Eq. (\ref{eq:RdRp}) where $ \Phi = n a^2 \left( 1- \sqrt{1-e^2} \right) \approx n a^2 e^2 / 2$ is the action conjugate to the angle $\Delta \varpi \equiv \varpi - \varpi_p$. As a result, we obtain the following expression:
\begin{equation}
    R = n a^2 \left[ \frac{1}{2} \left( A - A_{d,p} \right) e^2 + B_p e \cos \Delta\varpi \right] .
    \label{eq:Rgeneral}
\end{equation} 
This completes our development of the disturbing function.

\begin{figure}[t!] 
\epsscale{1.1}
\plotone{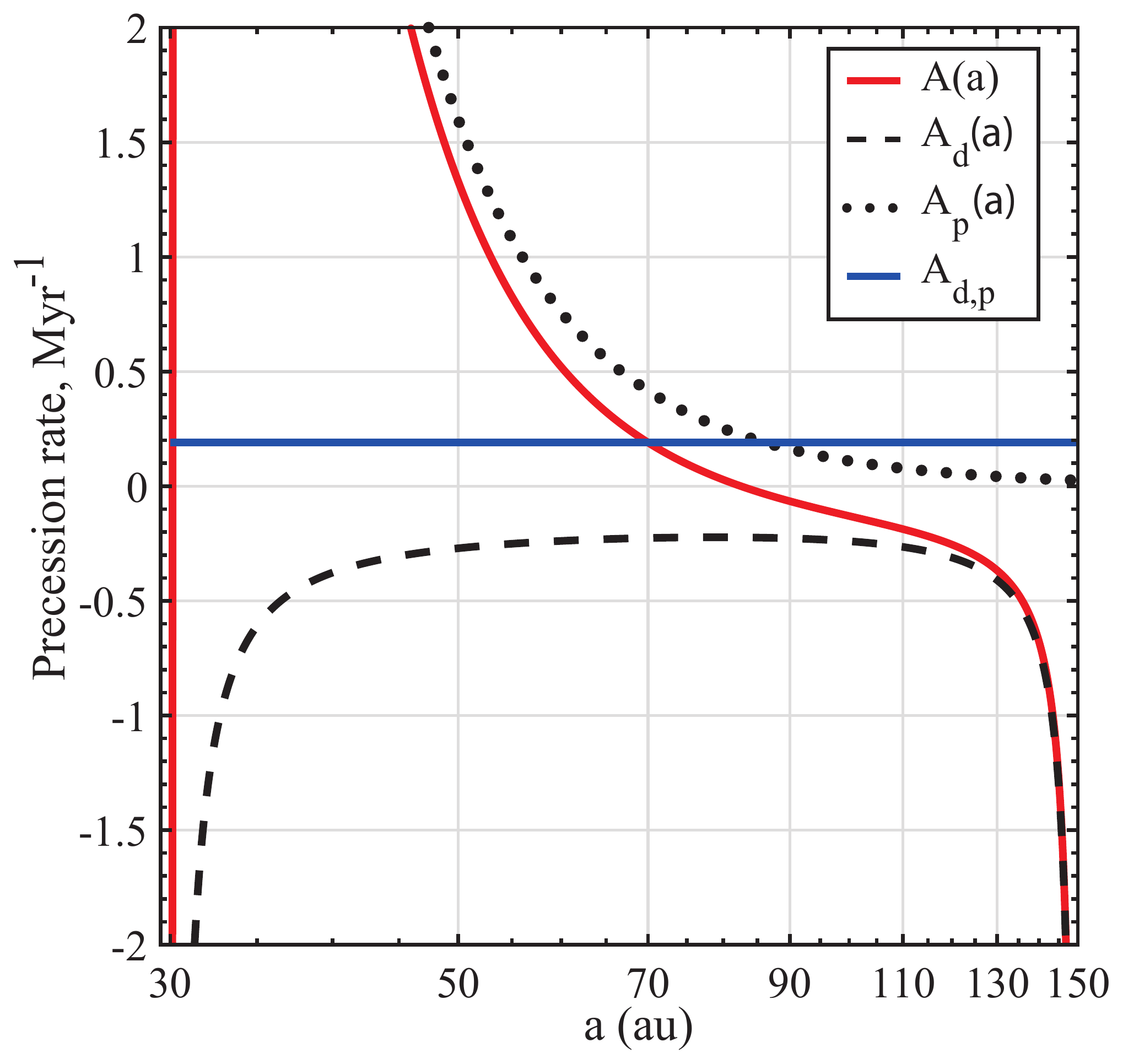}
\caption{Planetesimal free precession rate $A = A_d + A_p$ due to both the planet and the disk as a function of semimajor axis (red curve). Dotted and dashed curves represent $A_p(a)$ and $A_d(a)$, respectively. The blue line represents the rate of planetary precession $A_{d,p}$ due to the disk. Calculations assume a $20 M_{\earth}$ disk with $p=1$ extending from $a\inn=30$ au to $a\out=150$ au, and a $0.6 M_J$ planet at $a_p = 20$ au around a $1.09 M_{\odot}$ star (\texttt{Model A}, Table \ref{table:models}). Note that  $A(a) = A_{d,p}$ at two locations: at $70$ au and at $\simeq a\inn$.
\label{fig:A_a_figure}}
\end{figure}

Note that for the particular set of parameters in equations (\ref{eq:Aplanet}), (\ref{eq:Adisc}), (\ref{eq:Adp}), the planetesimal free precession rate $A$ at $a =  70$ au is comparable to that of the planetary orbit, $A_{d,p}$. In Figure \ref{fig:A_a_figure} we show the radial behavior of $A = A_d + A_p$,  together with the curve for $A_{d,p}$. The fact that $A(a) = A_{d,p}$ at certain semimajor axes has very important implications for planetesimal dynamics; see Section \ref{subsec:resonance}.

\subsection{Evolution equations and their solution}
\label{subsec:evol_eq_theory}

The secular evolution of a planetesimal orbit in the combined potential of the planet and the disk can be determined by Lagrange's planetary equations \citep{mur99}. Introducing the eccentricity vector $\textbf{e} = (K, H) = e (\cos \Delta\varpi, \sin \Delta\varpi)$,  convenient for describing the dynamics in the frame corotating with the planet \citep[e.g.][]{hep80}, we find that:
\begin{eqnarray}
    \frac{dK}{dt}&\approx&  \frac{- 1}{n a^2} \frac{\partial R}{\partial H}  =    - (A - A_{d,p} ) H    ,
\nonumber
\\
    \frac{dH}{dt} &\approx&  \frac{1}{n a^2} \frac{\partial R}{\partial K}  =   (A - A_{d,p}) K + B_p  .
\label{eq:EOM_analytical}
\end{eqnarray}
Note that in the case of a massless disk ($A_{d,p} = 0, A=A_p$), one recovers the evolution equations due to a non-precessing perturbing planet \citep[e.g.][]{mur99}.

The system of equations (\ref{eq:EOM_analytical}) admits a general solution given by the superposition of the `free' and `forced' eccentricity vectors, $\mathbf{e}(t) = \mathbf{e}_{\rm free}(t) + \mathbf{e}_{\rm forced}(t)$ \citep{mur99}. In particular, when planetesimals are initiated on circular orbits, $K(0)= H(0) =0$, we have $e_{\rm free} = e_{\rm forced}$ and the evolution of planetesimal orbits is described by:
\begin{eqnarray}
    e(t) &=&  2 \bigg| e_{\rm forced} \sin \left( \frac{A - A_{d,p}}{2} t \right) \bigg| ,
    \label{eq:e_t_solution}
    \\
    \tan \Delta\varpi(t) &=& \tan \left( \frac{A - A_{d,p}}{2}t - \frac{\pi}{2} \right) ,
    \label{eq:w_t_solution}
\end{eqnarray}
where $\Delta \varpi$ stays in the range $[-\pi, \pi]$, and the forced eccentricity is given by
\begin{equation}
    e_{\rm forced}(a) 
    =   \frac{-B_p(a)}{A(a) - A_{d,p}} 
    =   \frac{-B_p(a)}{A_d(a) + A_p(a) - A_{d,p}} .
    \label{eq:eforced}
\end{equation}
Equations (\ref{eq:e_t_solution})--(\ref{eq:eforced}) represent the key solutions needed for our work. We remark that this framework has been previously verified against direct orbit integrations of test particles in disks \citep[e.g.][]{sil15, fontana, irina18}.

For illustrative purposes, in Figure \ref{fig:simA_e_w_a_time} we show the radial profiles of instantaneous eccentricities (left panels) and longitudes of pericenter (relative to the planet, right panels) of planetesimals computed using Eqs. (\ref{eq:e_t_solution}) and (\ref{eq:w_t_solution}) (i.e. for $e(0) = 0$) at different times, as indicated in each panel. The calculations assume the same disk-planet parameters as in Fig. \ref{fig:A_a_figure} and we have taken  $e_p = 0.05$ -- the parameters of the fiducial disk-planet model (\texttt{Model A}, Table \ref{table:models}) which we consider in details later in this work (Section \ref{sec:results_new}). Furthermore, here we have sampled secular evolution using $N = 5000$ planetesimals with semimajor axes distributed logarithmically between $a\inn$ and $a\out$, i.e. with a ratio of spacing $\beta = (a\out/a\inn)^{1/N} \approx 1.0003$, each of which is represented by a blue dot in Fig. \ref{fig:simA_e_w_a_time}. We note that, as is typical for secular evolution, the eccentricity oscillation at a given semimajor axis is bounded between the initial value of $0$ and $e_m(a) = 2|{e}_{\rm forced}(a)|$ (the red lines in left panels of Fig. \ref{fig:simA_e_w_a_time}). Moreover, as expected, the period of each eccentricity oscillation in the frame corotating with the planet is given by $\tau_{\rm sec} = 2\pi/(A-A_{d,p})$.

\subsection{Planetesimal eccentricity behavior and secular resonances} \label{subsec:resonance}

We now describe the essential features of planetesimal dynamics in the combined disk-planet potential\footnote{For detailed summary of the dynamics in an analogous setup (in application to planetesimal dynamics in circumbinary disks), see \citet{RRptype} and \citet{silsbeekepler}.}. In general, planetesimal orbits evolve differently depending on their free precession rate $A(a)$ relative to that of the planet $A_{d,p}$, i.e. for $A(a) > A_{d,p}$ or $A(a) < A_{d,p}$ -- see Eqs. (\ref{eq:e_t_solution}), (\ref{eq:w_t_solution}).

\begin{figure}[t!]
\epsscale{1.20}
\plotone{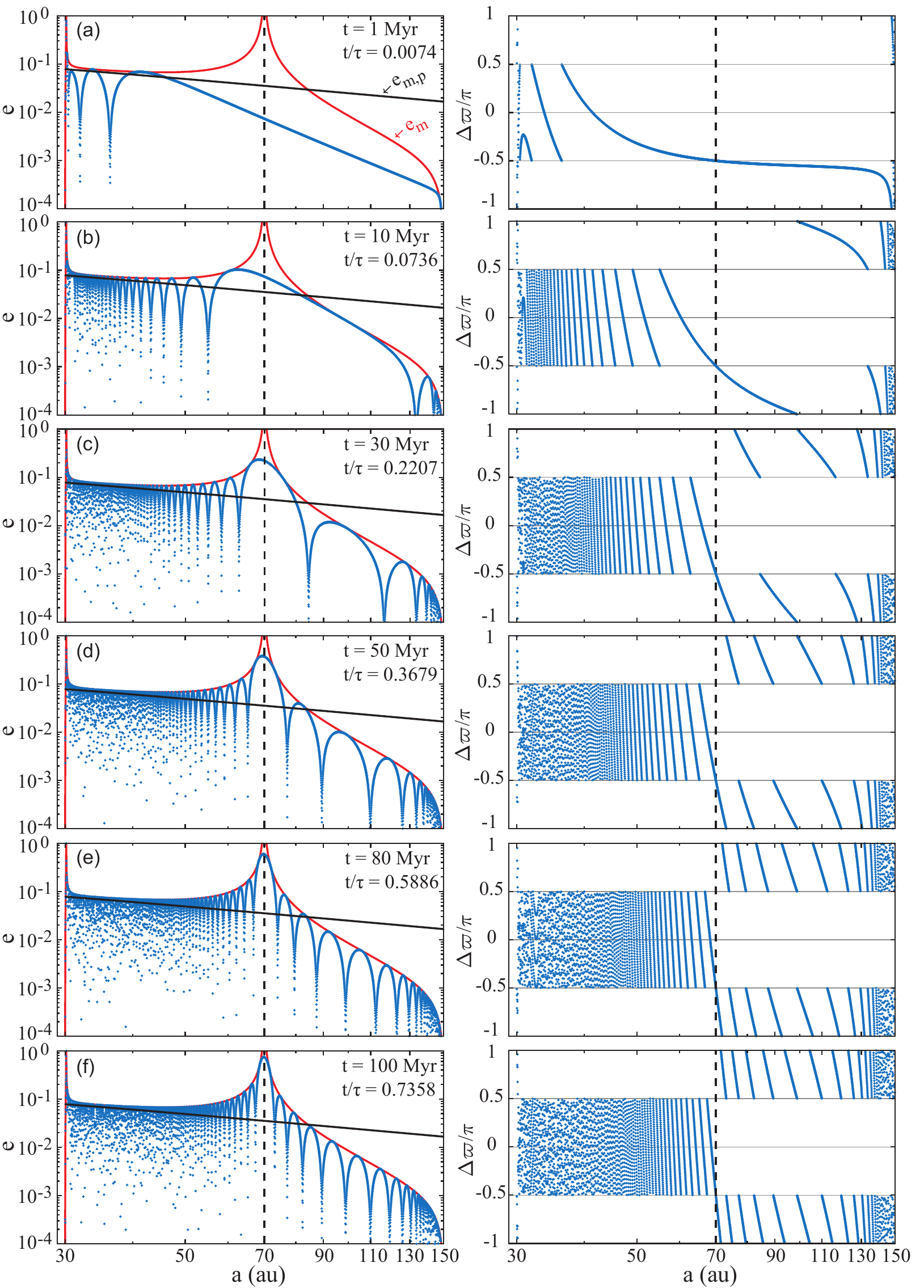}
\caption{Snapshots of the planetesimal eccentricities $e$ (left panels) and apsidal angles $\Delta\varpi$ (right panels, measured relative to that of the precessing planet) as a function of semimajor axis $a$ after $t = 1, 10, 30, 50, 80$ and $100$ Myr of evolution (top to bottom). The time is also indicated relative to $\tau \approx 135$ Myr,  Eq. (\ref{eq:axi_timescale}). The planetesimals were initiated on circular orbits in the fiducial disk-planet model (\texttt{Model A}, Table \ref{table:models}). The maximum of eccentricity oscillations $e_m = 2|e_{\rm forced}|$ (Eq. \ref{eq:eforced}) is shown by the red lines. For reference, the solid black lines show the maximum planetesimal eccentricities driven by the planet in the absence of the disk ($e_{m,p}$, Eq. \ref{eq:eforcedplanet}). The dashed vertical lines show the secular resonance location ($a\res = 70$ au), where eccentricities diverge in the course of evolution. One can clearly see that at the resonance $\Delta\varpi = -\pi/2$ at all times. Note also the resonance near the disk inner edge. This figure is available as an animation in the electronic edition of the journal. The animation runs from $t=0$ to $t = \tau \approx 135$ Myr with a duration of $36$ seconds.
\label{fig:simA_e_w_a_time}}
\end{figure}

For the particular set of parameters in Figs. \ref{fig:A_a_figure} and \ref{fig:simA_e_w_a_time}, we see that the regime $A(a)>A_{d,p}$ is realized at small separations from the planet, where the precession rate of planetesimals is dominated by the planet so that $A \approx A_p$ (except near $a\inn$ where $A_d$ diverges due to disk edge effects, \citet{sil15}); see also Eqs. (\ref{eq:Aplanet}),  (\ref{eq:Adisc}). In this \textit{planet-dominated} regime planetesimal orbits precess in the same direction as the planet (i.e. prograde, see Eq. \ref{eq:w_t_solution} and right panels of Fig. \ref{fig:simA_e_w_a_time}), and we have ${e}_{\rm forced} > 0$ (Eq. \ref{eq:eforced}). Thus, as planetesimal orbits evolve, the apsidal angles $\Delta\varpi$ remain constrained within $[-\pi/2, \pi/2]$ at all times. Moreover, planetesimals attain their maximum eccentricity when their orbits are aligned with that of the planet, i.e. when $\Delta \varpi = 0$; see Eq. (\ref{eq:e_t_solution}) and Fig. \ref{fig:simA_e_w_a_time}.  Assuming $A_p \gtrsim A_{d,p}$, the maximum planetesimal eccentricity in this regime is  $e_{m, p} \approx |2 e_{\rm forced, p}|$ with \citep[e.g.][]{mur99}
\begin{eqnarray}
   e_{\rm forced, p}  = \frac{-B_p}{A_p}
    = \frac{b_{3/2}^{(2)}(a_p/a)}{b_{3/2}^{(1)}(a_p/a)} e_p
    &\approx&  \frac{5}{4} \frac{a_p}{a} e_p , 
    \label{eq:eforcedplanet} 
    \\
    &    \approx  & 1.8 \times 10^{-2} \frac{a_{p,20}}{a_{70}} \frac{e_p}{0.05}  , 
    \nonumber
\end{eqnarray}
see Eq. (\ref{eq:eforced}), where we have used the approximations $b_{3/2}^{(1)}(\alpha) \approx 3\alpha$ and $b_{3/2}^{(2)}(\alpha) \approx (15/4)\alpha^2$ valid for small $\alpha$. This is the limit of a \textit{massless} disk, a configuration most often adopted in studies of debris disks. In the course of evolution, planetesimals in this regime will form an eccentric structure largely aligned with the planetary orbit \citep[e.g.][]{wyattetal99}.

In the opposite \textit{disk-dominated} limit, far from the planet (and for $a\approx a\inn$, which we discuss later), Figure \ref{fig:A_a_figure} shows that the precession rate of planetesimals is dominated by the disk so that $A \approx - |A_d| \lesssim A_{d,p}$. In this regime planetesimal orbits undergo retrograde free precession (see Eq. \ref{eq:w_t_solution} and right panels of Fig. \ref{fig:simA_e_w_a_time}), and we have $e_{\rm forced} < 0$. Thus, the apsidal angles $\Delta\varpi$ are confined within the range $\pm[\pi/2, \pi]$ at all times. Moreover, planetesimals attain their maximum eccentricity when their orbits are \textit{anti-aligned} with the planetary orbit, i.e. when $|\Delta \varpi| = \pi$; see Eq. (\ref{eq:e_t_solution}). Assuming $A_{d,p} \rightarrow 0$ for simplicity, the maximum eccentricity in this regime is $e_{m,d} \approx |2 e_{\rm forced, d}|$ with
\begin{eqnarray}
    |e_{\rm forced, d}| = \bigg| \frac{B_p}{A_d} \bigg| 
    &\approx &  \frac{15 e_p}{16 |(2-p)\psi_1|} \frac{m_p}{M_d} \left(\frac{a_p}{a}\right)^3
    \left( \frac{a\out}{a} \right)^{2-p} ,
    \label{eq:eforced_disk}
    \\
    &\approx& 4.7 \times 10^{-3} \frac{m_p}{M_d} \frac{e_p}{0.05} \frac{a_{p,20}^3 a_{\rm out, 150}}{a_{70}^4}     ,
    \nonumber
\end{eqnarray}
where the numerical estimate assumes $p=1$ and $a\inn \ll a \ll a\out$ so that $\psi_1 \approx -0.5$. Equation (\ref{eq:eforced_disk}) shows that planetesimal eccentricities in the disk-dominated regime decline more rapidly with $a$ than in the planet-dominated regime, and their magnitude is suppressed -- an effect pointed out in \citet{RRptype}. In the course of evolution, planetesimals in this regime will form an eccentric structure anti-aligned with the planetary orbit.

\subsubsection{Main secular resonance}

More importantly, one can clearly see that the transition between planet- and disk-dominated regimes occurs via a \textit{secular eccentricity resonance} where $A(a) = A_{d,p}$; see Fig. \ref{fig:A_a_figure} \citep[see also][]{RRptype, silsbeekepler}. This resonance emerges because the relative precession between the planetesimal orbits and the planetary orbit vanishes, while the torque exerted by the non-axisymmetric component of the planet is non-zero. At and around the locations of secular resonances, $a = a\res$, planetesimal eccentricities are forced to arbitrarily large values (in linear approximation), see left panels of Fig. \ref{fig:simA_e_w_a_time}. This is because the denominator in Eq. (\ref{eq:eforced}) becomes small,  introducing a singularity into the secular solution\footnote{Including higher order terms (in eccentricities)  of the disturbing function (\ref{eq:Rgeneral}) imposes a finite upper limit on the amplitude of $e_{\rm forced}$ at secular resonance \citep{malhotra98, wardhahn1998}.} \citep{RRptype}. By taking a limit $A(a\res) \rightarrow A_{d,p}$ in Eq. (\ref{eq:eforced}) we find that the growth of eccentricity at the resonance occurs linearly in time, $e(t) = t / \tau$, with a characteristic timescale given by
\begin{equation}
  \tau = \frac{1}{\left| B_p(a_{\rm res}) \right| } \approx  158 ~{\rm Myr} \frac{ 0.6 M_{J}  }{ m_p  }  \frac{0.05}{e_p} \frac{a_{\rm res, 70}^{9/2}}{ a_{p, 20}^3 } M_{c,1.09}^{1/2}     ,
  \label{eq:axi_timescale}
\end{equation}
where the approximation is valid for $a_p \ll a\res$. Eq. (\ref{eq:axi_timescale}) also explains why the eccentricities at the resonance near the disk inner edge are pumped up more quickly than at the resonance at $70$ au, see left panels of Fig. \ref{fig:simA_e_w_a_time}.

Moreover, we can see from the right panels of Fig. \ref{fig:simA_e_w_a_time} that at the resonance $\Delta\varpi$ remains fixed at $-\pi/2$, as expected from Eq. (\ref{eq:w_t_solution}). In Section \ref{sec:res_location} we will show that such secular resonances are generic: they occur for a large range of disk-to-planet mass ratios, $ 10^{-4} \lesssim M_d/m_p \lesssim 2$, for all $a_p \lesssim a\inn$.

\begin{figure}[t!]  
\epsscale{1.05}
\plotone{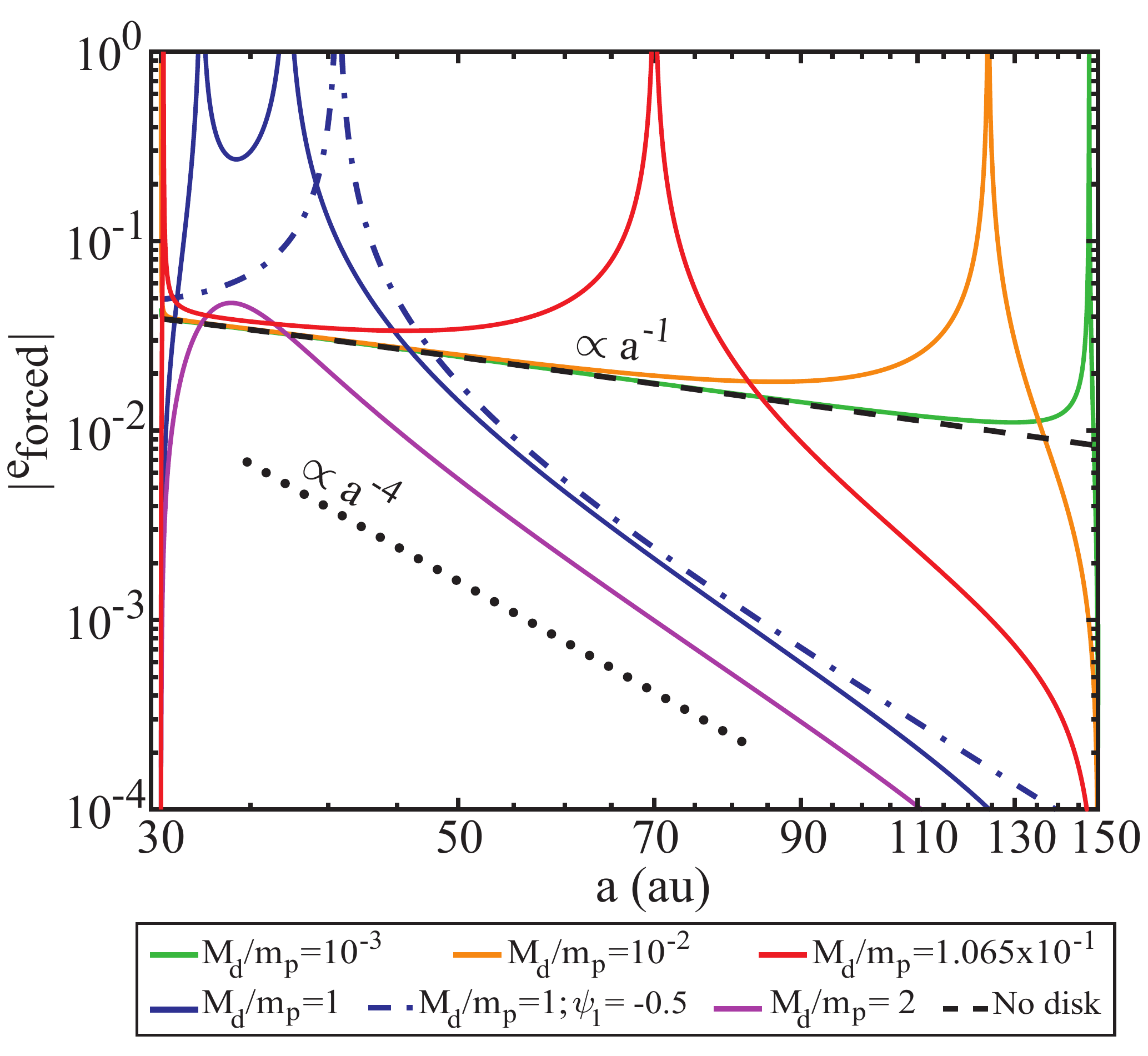}
\caption{Forced eccentricities of planetesimals as a function of their semimajor axis $a$, computed for different values of $M_d/m_p$ (with fixed $m_p = 0.6 M_J$). The calculations assume all other system parameters are as in Figs. \ref{fig:A_a_figure}, \ref{fig:simA_e_w_a_time}. All of these curves scale linearly with the planetary eccentricity $e_p$, which we have taken to be $0.05$ in this calculation. For reference, the black dashed line shows forced eccentricity in the case of a massless disk $e_{\rm forced, p}$ (Eq. \ref{eq:eforcedplanet}), and the dotted line illustrates the asymptotic behavior of eccentricity given by $e_{\rm forced, d}$ (Eq. \ref{eq:eforced_disk}). Note the occurrence of two secular resonances for $10^{-3} \leq M_d/m_p \leq 1$, with one of them being near the inner disk edge. See text (\S \ref{subsec:resonance}) for details. 
\label{fig:eforced_a_figure}}
\end{figure}

To further illustrate the analysis above, Figure \ref{fig:eforced_a_figure} shows the radial profiles of planetesimal forced eccentricities computed for different values of disk mass. The calculations are done for the same planetary parameters as in Figs. \ref{fig:A_a_figure}, \ref{fig:simA_e_w_a_time}. The most pronounced feature in Fig. \ref{fig:eforced_a_figure} is the occurrence of a secular resonance within the disk (apart from the one very close to $a\inn$, see below) for $ 10^{-3} \leq M_d/m_p  \leq 1$, where $e_{\rm forced}$ diverges. At the same time, $e_{\rm forced}$ asymptotically approaches $e_{\rm forced, p}$ inward of the resonance, i.e. where $A \gtrsim A_{d,p}$, whereas $e_{\rm forced} \rightarrow e_{\rm forced, d}$ external to it, i.e. where $A \lesssim A_{d,p}$ (which is, of course, possible only if $a\inn \lesssim a\res \lesssim a\out$).  At the highest disk mass, $M_d/m_p = 2$, there are no secular resonances as the disk dominates planetesimal precession throughout the whole disk.

We note that in the region where the dynamics is dominated by the disk $e_{\rm forced}(a)$ does not follow the simple power law profile $\propto a^{-4}$ given by Eq. (\ref{eq:eforced_disk}). By and large, this is because the disk edge effects neglected in computing Eq. (\ref{eq:eforced_disk}) render $\psi_1 = \psi_1(a)$ in a non-trivial manner, even when $a_{\rm in} \lesssim a \lesssim a\out$ \citep{sil15}. For instance, it is evident in Fig. \ref{fig:A_a_figure} that $A_d(a)$ behaves more like a constant for $a\inn \ll a \ll a\out$ rather than as $A_d \propto a^{-1/2}$ (Eq. \ref{eq:Adisc}), implying that $|\psi_1| \propto a^{1/2}$ for the employed disk model. This will be important in \S  \ref{sec:res_location}. As a matter of fact, $\psi_1$ becomes independent of semimajor axis only in disks of infinite radial extent \citep{sil15}, whereas the radial range of our adopted disk is finite with $\delta = a\out/a\inn = 5$ (\S \ref{subsec:model_system})  .

\subsubsection{Secular resonance at $a_{\rm in}$}
\label{sec:SR_at_ain}

Finally, we clarify that the origin of the resonance at $\approx a\inn$ (apart from the one at $\gtrsim a\inn$) lies in the fact that $A_d \propto -|\psi_1|$ diverges as the sharp edges of a razor-thin disk are approached, see black dashed lines in Fig. \ref{fig:A_a_figure}. This makes $|A_d(a)| \sim A_p(a)$ as $a \rightarrow a\inn$, even for a modest value of disk mass. However, it is also known that disks with  $\Sigma_d$ dropping continuously near the edges rather than discontinuously, or disks with small but non-zero thickness, should exhibit finite $A_d$ near the edges \citep{irina18, SR19}; different from our disk model. Thus, in such more realistic disks, only a single resonance -- rather than two -- will occur. This is portrayed in Fig. \ref{fig:eforced_a_figure} for $M_d/m_p = 1$ by artificially stipulating $\psi_1(a) = -0.5$, i.e. by ignoring the edge effects \citep{sil15}.

\subsubsection{Secular resonances and gaps in debris disks}

To summarize, the analysis presented here elucidates that the disk gravity can have a considerable impact on the secular evolution of planetesimals. In the remainder of this paper, we exploit the feasibility of the discussed secular resonance as the basis of a mechanism for sculpting depleted regions, i.e. gaps, in debris disks.

The emergence of a gap could be understood as follows. Planetesimals on eccentric orbits spend most of their time near their apocenter, further away from their orbital semimajor axes. Thus, provided that a secular resonance occurs within the disk, we expect the surface density of planetesimals to be depleted around the resonance location where planetesimal eccentricities grow without bound. This reasoning, in essence, is similar to that presented by \citet{yelverton2018} where the authors show that two planets could carve a gap in an external \textit{massless} debris disk through their secular resonances. Additionally, given that generally planetesimals in the inner disk parts tend to apsidally align with the planet while those in the outer parts tend to anti-align, we expect the depleted region to have a non-axisymmetric shape. This effect has been previously pointed out by \citet{pearcewyatt15} in the context of secular interaction  between a debris disk and an interior, precessing planet.

\section{Characterization of Secular Resonances} 
\label{sec:resonances} 

We now investigate how the characteristics of the secular resonances -- i.e. their locations, their associated timescales for exciting eccentricities, and their widths -- depend on the properties of the disk and the planet. This will guide us in putting constraints on the possible disk-planet parameters that could reproduce the structure of an observed debris disk featuring a gap (\S \ref{sec:HD_constraints}).

\subsection{Location of secular resonances} 
\label{sec:res_location}

\begin{figure}[t!]
\epsscale{1.15}
\plotone{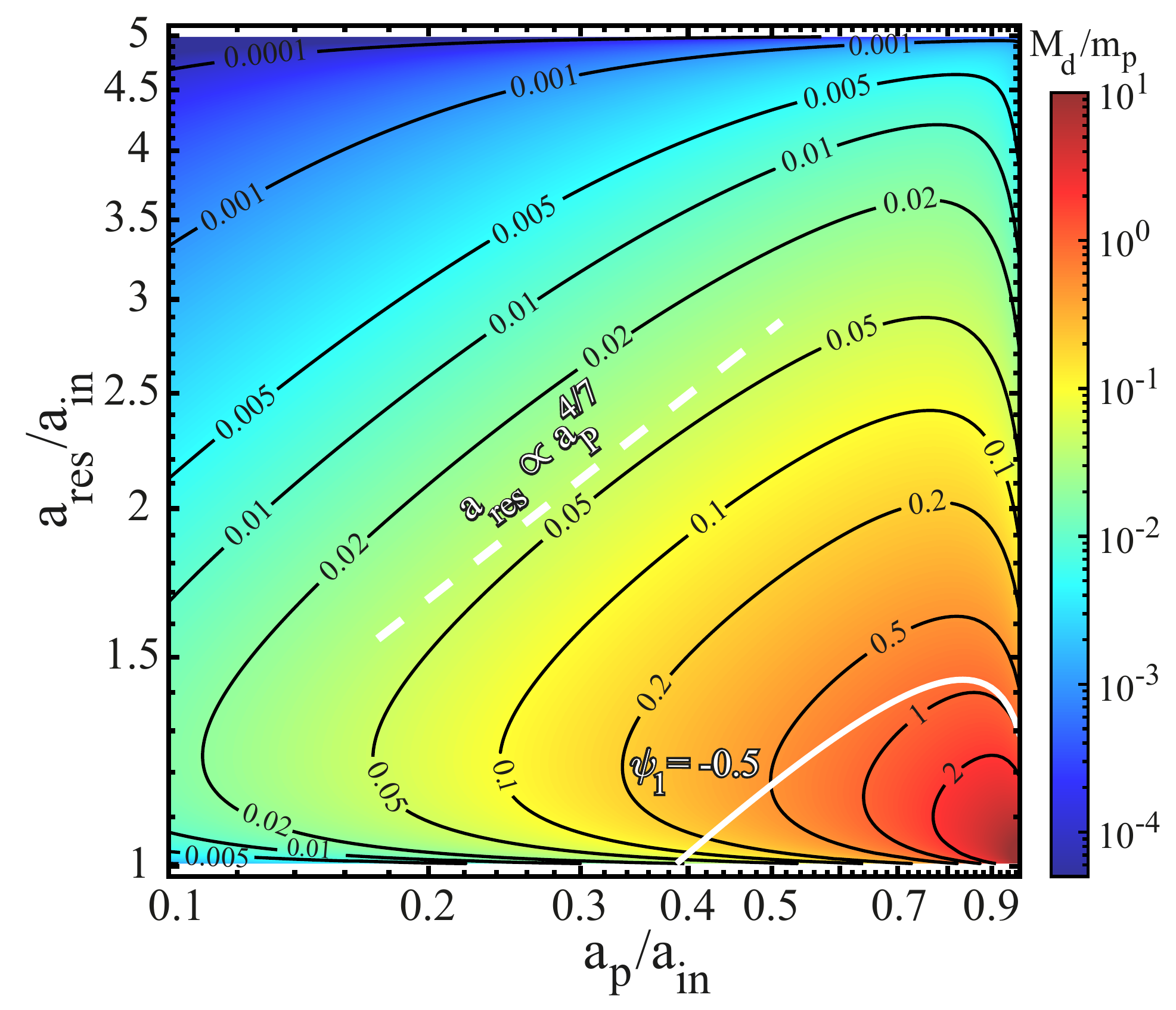}
\caption{Location of secular resonances relative to the disk inner edge $a\res/a\inn$ as functions of $a_p/a\inn$ and $M_d/m_p$. Calculations assume a power-law disk model with $p=1$ and $\delta \equiv a\out/a\inn = 5$. The full white line represents the contour for $M_d/m_p = 1$ obtained by ignoring disk edge effects, i.e. $\psi_1 = -0.5$. The dashed white line shows the scaling of $a\res$ with $a_p$ for fixed $M_d/m_p$, Eq. \ref{eq:num_scaling_mass}. See text (\S\ref{sec:res_location}) for details.
\label{fig:resonance_map}}
\end{figure}

As mentioned in Section \ref{subsec:resonance}, secular resonances occur at semimajor axes $a = a\res$ where the apsidal precession rates of both the planet and planetesimals are commensurate,
\begin{equation}
    A_d(a\res) + A_p(a\res) = \dot{\varpi}_p \equiv A_{d,p}  .
    \label{eq:resonance_condition}
\end{equation}
Using Equations (\ref{eq:Aplanet}), (\ref{eq:Adisc}) and (\ref{eq:Adp}), we can express the resonance condition (\ref{eq:resonance_condition}) in terms of the disk-to-planet mass ratio $M_d/m_p$ and the relevant semimajor axes, i.e. $a\res$, $a_p$, and $a\out$, scaled by $a\inn$:
\begin{eqnarray}
C_1 \psi_1  \frac{M_d}{m_p} \left( \frac{a_{\rm res}}{a_{\rm in}} \right)^{2-p}  &+& \frac{1}{4}  \frac{a_p}{a_{\rm res}} b_{3/2}^{(1)}\left( \frac{  a_p}{a\res} \right)
\nonumber\\
&  = &  
\frac{3}{4}  C_2 \phi_1^c   \frac{M_d}{m_p}  
\left( \frac{a_p}{a\inn}  \right)^3
\left( \frac{a_p}{a\res} \right)^{-3/2} . 
\label{eq:resonance_condition_literal}
\end{eqnarray}
Here $C_1 = (2-p)/(\delta^{2-p} -1)$ and $C_2 =  C_1 (1- \delta^{-p-1})/(p+1)$ are constants depending on the disk model. It follows from Eq. (\ref{eq:resonance_condition_literal}) that the locations of secular resonances can be computed \textit{relative} to the disk inner edge as functions of $a_p/a\inn$ and $M_d/m_p$. This is illustrated in Figure \ref{fig:resonance_map}, where we plot the contours of $M_d/m_p$ in the ($a_p/a\inn, ~ a\res/a\inn$) plane computed using our fiducial disk model i.e. $p=1$ and $\delta = 5$ (\S \ref{subsec:model_system}).

Figure \ref{fig:resonance_map} shows that for any given planet, two or no secular resonances occur within the disk provided that $10^{-4} \lesssim M_d/m_p \lesssim 2$. Additionally, we can see that for any $a_p/a\inn$ one of the resonances always occurs in the vicinity of the disk inner edge as described in \S \ref{sec:SR_at_ain}, i.e. $ a_{\rm res,1} \simeq a\inn$, and its location varies weakly with $M_d/m_p$. On the other hand, the second resonance occurs at semimajor axis $a_{\rm res,2} \gtrsim a_{\rm res,1}$ whose location changes significantly with varying $M_d/m_p$. Indeed, with increasing $M_d/m_p$ (at fixed $a_p/a\inn$) this resonance is pushed inwards from $ \simeq a\out$ towards the inner resonance at $\simeq a\inn$ until both resonances `merge', i.e. the distance between them approaches zero. Figure \ref{fig:eforced_a_figure} provides a complementary view of this behavior. Looking at Fig. \ref{fig:resonance_map} we also see that, for planets closer to the disk,  larger $M_d/m_p$ is necessary to maintain the resonance at a given semimajor axis.

We recall that the existence of the inner resonance is mainly due to the disk edge effects. That is, the divergence of $A_d(a) \propto -|\psi_1(a)|$ as $a\rightarrow a\inn$ allows the resonance condition (\ref{eq:resonance_condition}) to be satisfied around $\approx a\inn$, even for relatively small values of $M_d$ (Section \ref{subsec:resonance}). This explains why for a given $a_p/a\inn$ the resonance at $a_{\res,1}$ is constrained to be very close to $\simeq a\inn$ irrespective of $M_d/m_p$. In the absence of edge effects this inner resonance will not exist, resulting in a single resonance for fixed system parameters rather than two. This is illustrated in Fig. \ref{fig:resonance_map} for $M_d/m_p=1$ by setting $\psi_1(a) = -0.5$ (white full line).

The behavior of the resonance locations can be explained analytically. Consider the approximate form of the resonance condition, Eq. (\ref{eq:resonance_condition}), in the limit of $a_p/a\inn \rightarrow 0$ so that $A_{d,p}$ is negligible and one can use the asymptotic limit of $b_{3/2}^{(1)}$, and the two terms on the left hand side of Eq. (\ref{eq:resonance_condition_literal}) balance each other (recall that $\psi_1 <0$). It is then easy to demonstrate that for a resonance to occur at $a_{\rm in} \lesssim a\res \lesssim a\out$, the disk mass must be given by
\begin{eqnarray}
\frac{M_d}{m_p} & \approx & \frac{3\delta^{2-p}}{4 | (2-p) \psi_1(a\res) |} \left( \frac{a_p}{a\inn} \right)^2 \left( \frac{a\res}{a\inn} \right)^{p-4} ,
\label{eq:num_scaling_mass}
\\
&\approx & 0.15 ~ a_{p, 20}^2 ~   a_{\rm res, 70}^{-3.5}  ,
\nonumber
\end{eqnarray}
where the numerical estimate is obtained for our fiducial disk model ($p=1$, $\delta =5$), for which $|\psi_1(a)| \propto a^{1/2}$ when $a\inn \ll a \ll a\out$, see Section \ref{subsec:resonance} \footnote{In an infinitely extending disk, i.e. as $\delta\rightarrow \infty$, $\psi_1$ becomes independent of semimajor axis  e.g. $\psi_1(a) = -0.5$ for $p=1$. In this case, Eq. (\ref{eq:num_scaling_mass}) would read as $M_d/m_p \approx 0.26 ~ a_{p,20}^2 ~ a_{\res, 70}^{-3}$.}. Fixing $M_d/m_p$ in Eq. (\ref{eq:num_scaling_mass}) then approximates the slopes of the contours in Fig. \ref{fig:resonance_map} reasonably well -- see the white dashed line. As expected, the numerical results deviate from the scaling in Eq. (\ref{eq:num_scaling_mass}) both as $a\res \rightarrow a\inn$ or $a\out$, where $\psi_1$ diverges, and as $a_p \rightarrow a\inn$, since $A_{d,p}$ becomes non-negligible.

\subsection{Timescale for eccentricity excitation} \label{sec:res_timescale}

We now consider how the eccentricity excitation timescale varies as a function of model parameters. To this end, we make use of the definition of $\tau$ given by Eq. (\ref{eq:axi_timescale}), which quantifies the time it takes for initially circular orbits to reach $e=1$ at the resonance. We note that $\tau$ is a strong function of the resonance location, and it explicitly depends on the parameters of the planet but not the disk. This is because the disk, assumed to be axisymmetric in our model (Section \ref{sec:theory}), does not contribute to eccentricity excitation.

\begin{figure}[t!]
\epsscale{1.15}
\plotone{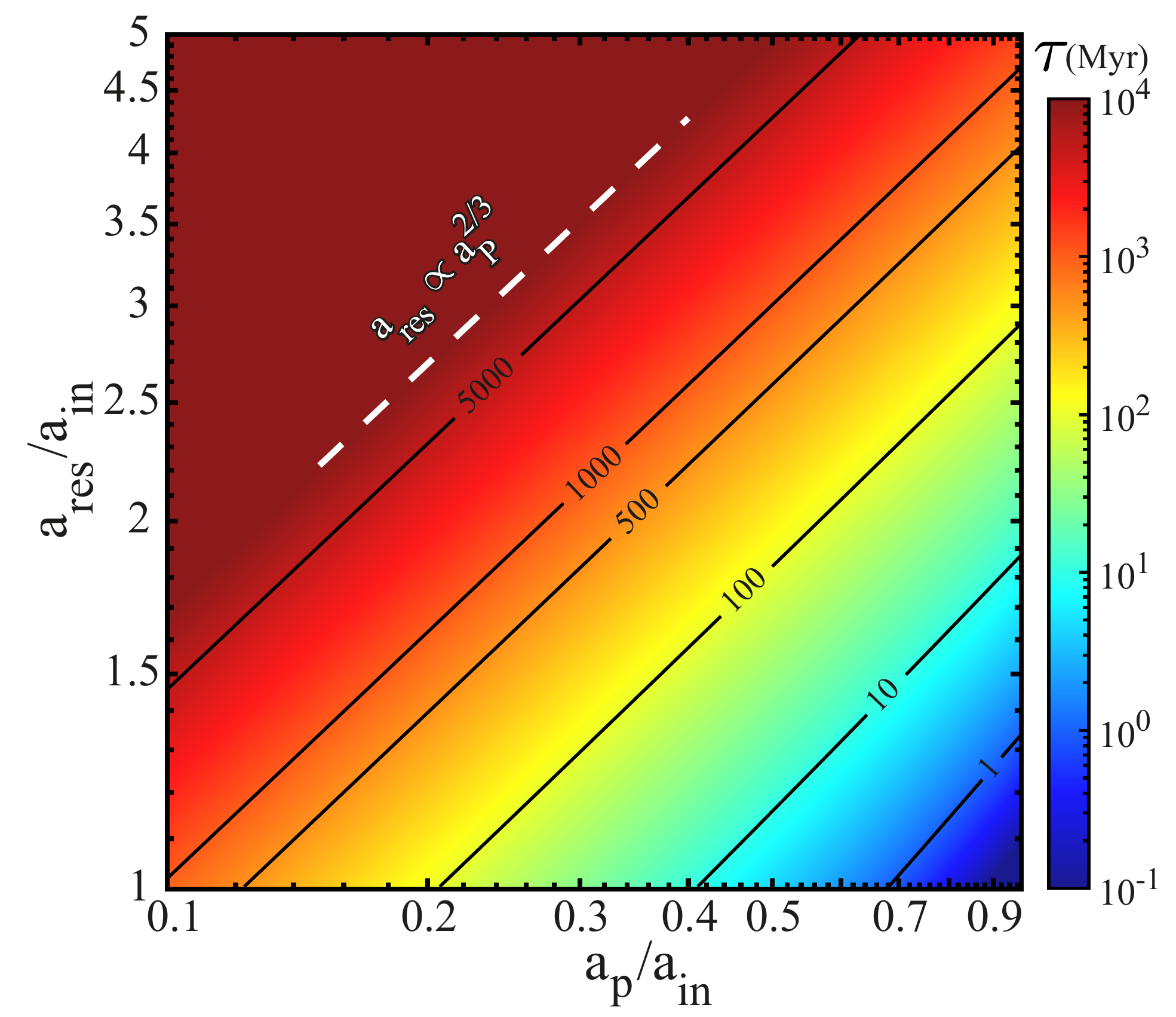}
\caption{Contour plot of the timescale $\tau$  for exciting planetesimal eccentricities by the secular resonance (Eq. \ref{eq:axi_timescale}), in the space of $a_p/a\inn$ and $a\res/a\inn$. The calculations assume a planet with $m_p = 100 M_{\earth}$ and $e_p = 0.1$ around a solar-mass star. The white dashed line shows the scaling of $a\res$ with $a_p$ for a fixed value of $\tau$. See text (\S \ref{sec:res_timescale}) for details.
\label{fig:timescale_map}}
\end{figure}

In Figure \ref{fig:timescale_map} we plot the contours of $\tau$ in the $(a_p/a\inn,~  a\res/a\inn)$ plane for a particular choice of planetary mass and eccentricity,  $m_p = 100 M_{\earth}$ and $e_p = 0.1$, assuming a solar-mass star. It is evident that the timescales are shorter when the planet and the resonance location are closer together, i.e. in the lower-right corner of parameter space where $a\res/a_p \rightarrow 1$. Note that for the adopted planetary parameters, over a broad range of parameter space the timescales range from $\sim 10$ Myr to few Gyr; this is comparable to the ages of observed debris disks. Moreover, the slopes of the contours in Fig. \ref{fig:timescale_map} can be explained by setting $\tau$ to a constant in Eq. (\ref{eq:axi_timescale}): this yields the scaling $a\res \propto a_p^{2/3}$ illustrated by the white dashed-line in Fig. \ref{fig:timescale_map}.

Finally, Equation (\ref{eq:axi_timescale}) shows that $\tau$ is inversely proportional to both the planetary mass and eccentricity. Thus, more massive or eccentric planets exert larger torque and excite planetesimal eccentricities more quickly, shortening the timescale $\tau$ when $a_p/a\inn$ and $a\res/a\inn$ are kept fixed. This means that in Figure \ref{fig:timescale_map} the contours of $\tau$ will be shifted to the left (right) when the product of $m_p$ and $e_p$ is increased (decreased).

\subsection{Resonance width} \label{sec:res_width}

We now quantify the range of semimajor axes $w$ over which resonances act to  significantly excite planetesimal eccentricities. To this end, we  follow\footnote{For an alternative method, see \citet{levisonagnor}.} \citet{yelverton2018} and calculate the distance over which the forced planetesimal eccentricities $e_{\rm forced}(a)$ exceed a \textit{constant threshold value} $\tilde{e}$. That is, we define $w$ as the difference (in absolute values) between the two values of semimajor axis $a_i$ ($i = 1, ~2$) satisfying
\begin{equation}
    \tilde{e} = | e_{\rm forced}(a_i) | = \bigg| \frac{- B_p(a_i)}{A(a_i) - A_{d,p}}  \bigg|
\label{eq:width_numeric}
\end{equation}
in the vicinity of a given resonance. Here, we clarify that this definition serves as a proxy for the significance of a given resonance, and it does not necessarily correspond to the actual widths of gaps that we expect to observe\footnote{This is not least because the actual widths of gaps depend non-trivially on the spatial distribution of planetesimals, i.e. the profiles (and gradients) of both $e(a)$ and $\varpi(a)$ \citep{statler01}.}.

In Equation (\ref{eq:width_numeric}), the planetary and disk masses appear only through their ratio $M_d/m_p$, and the two relevant semimajor axes -- $a_i$ and $a_p$ -- could be expressed relative to $a\inn$; see Eqs. (\ref{eq:Aplanet}) -- (\ref{eq:Adp}). Furthermore, the ratio $M_d/m_p$ could be related to $a_p/a\inn$ and $a\res/a\inn$ by using the condition for secular resonance, Eqs. (\ref{eq:resonance_condition}), (\ref{eq:resonance_condition_literal}). Thus, we can compute the resonance width $w$ \textit{relative} to $a\inn$ as functions of $a_p/a\inn$ and $a\res/a\inn$ only, once $\tilde{e}$ and $e_p$ are specified (recall that $B_p \propto e_p$, Eq. \ref{eq:Bplanet}).

The threshold eccentricity $\tilde{e}$ in Eq. (\ref{eq:width_numeric}) represents an ad hoc parameter, necessitating a physical justification for a particular choice of its value. To this end, we note that the presence of a physical gap within the disk is subject to the condition that planetesimal eccentricities are larger around the resonances than elsewhere. Away from the resonances, the forced planetesimal eccentricity is maximized near the disk inner edge where, approximately, $e_{\rm forced}(a\inn) \rightarrow e_{\rm forced, p}(a\inn)$ which can not exceed $e_p$; see Eq. (\ref{eq:eforcedplanet}), Fig. \ref{fig:eforced_a_figure}.  Based on this reasoning we adopt $\tilde{e} = e_p$ in what follows, unless stated otherwise.

\begin{figure}[t!]
\epsscale{1.15}
\plotone{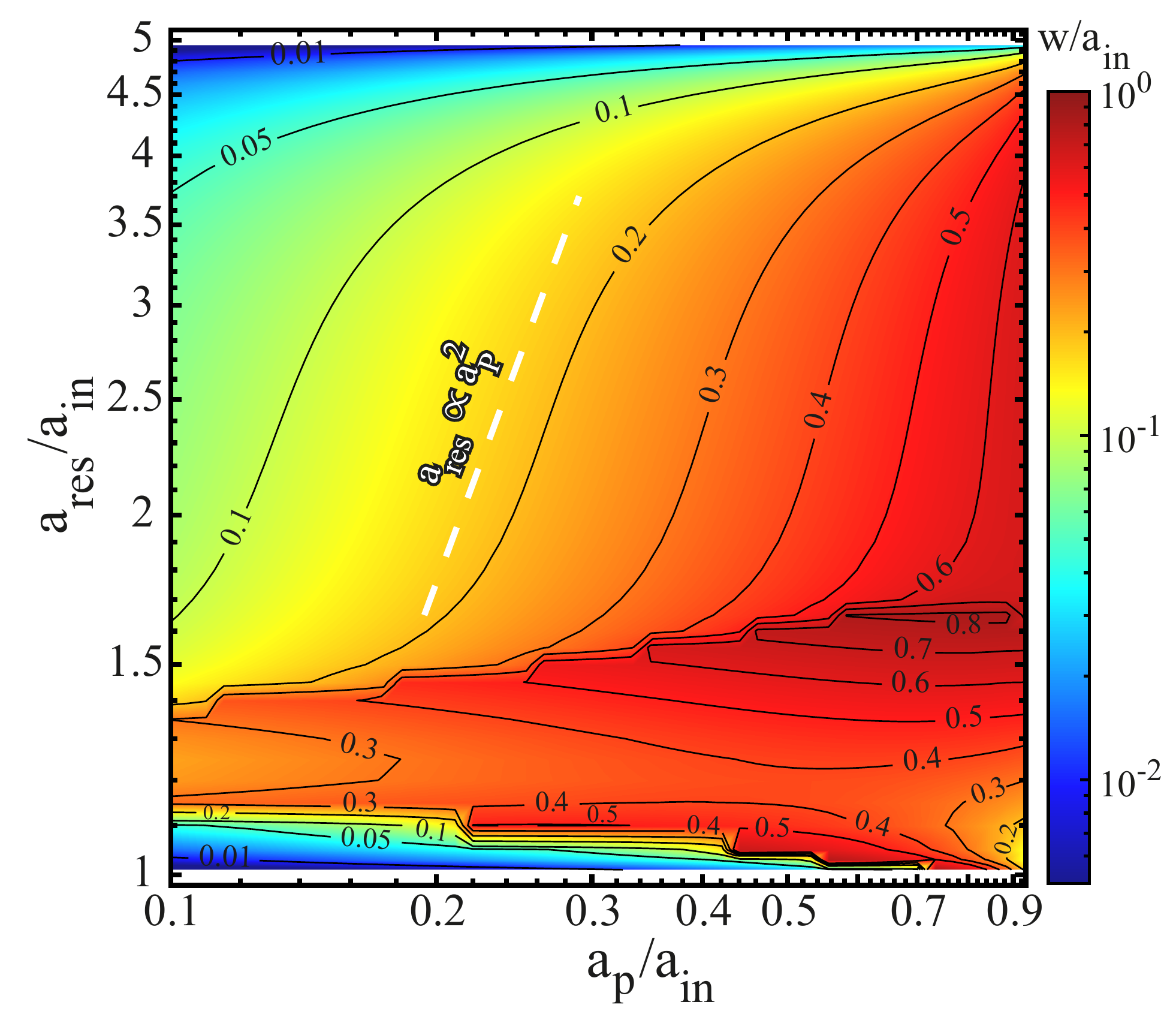}
\caption{Contour plot of the resonance width $w$ relative to $a\inn$ (Eq. \ref{eq:width_numeric}) in the space of $a_p/a\inn$ and $a\res/a\inn$, computed using $\tilde{e} = e_p$ and the same disk parameters as in  Fig. \ref{fig:resonance_map}. The white dashed line shows the scaling of $a\res$ with $a_p$ for a fixed value of $w/a\inn$, Eq. (\ref{eq:width_scaling}). See text (\S \ref{sec:res_width}) for details.
\label{fig:width_map}}
\end{figure}

In Figure \ref{fig:width_map} we plot the contours of $w/a\inn$ in the $(a_p/a\inn, ~a\res/a\inn$) plane for our fiducial disk model  with $p=1$ and $\delta = 5$ (see \S\ref{subsec:model_system}),  assuming $\tilde{e} = e_p$. Looking at Figure \ref{fig:width_map}, we see that increasing the planetary semimajor axis for a fixed $a_{\rm in}$ tends to generally broaden the width of a given resonance. This is, though, less obvious in the range $1.1 \lesssim a\res/a\inn \lesssim 1.5$  as the width there is a weaker function of $a_p/a\inn$.  Secondly (and relatedly), we see that for a given planetary semimajor axis, resonances occurring closer to the disk inner edge generally have larger widths compared to resonances further away; see also Fig. \ref{fig:eforced_a_figure}. The exception to this is if $a\res/a\inn  \simeq 1$, where the values of $w/a\inn$ are comparatively smaller, particularly in the lower-left corner of Fig. \ref{fig:width_map}.

To understand this behavior, we recall that for a given $a_p/a\inn$, our disk model with sharp edges has two resonance sites: one \textit{always} at $ a_{\rm res, 1} \simeq  a\inn $ and another further away at $a_{\rm res,1} \lesssim a_{\rm res,2} \lesssim a\out$; see Section \ref{sec:res_location}. In terms of Fig. \ref{fig:width_map}, this means that for a given $a_p/a\inn$ (and $M_d/m_p$, see Fig. \ref{fig:resonance_map}) if the resonances are well separated from each other, i.e. $a_{\rm res,1} \ll a_{\rm res,2}$, the inner resonance will be much narrower than the other. This behavior could be understood for instance by looking at the curves in Fig. \ref{fig:eforced_a_figure} for $M_d/m_p = 10^{-3}, 10^{-2}$ or $10^{-1}$, which show that the inner resonance width is insignificant.

On the other hand, for fixed ($a_p/a\inn, M_d/m_p$), if the resonances are close to each other such that $a_{\rm res,2}/a\inn \lesssim 1.5$ and $a_{\rm res,1} \simeq a\inn$ (see Fig. \ref{fig:resonance_map}), the resonances `merge' together yielding relatively large values of $w/a\inn$. What we mean by `merging' here is that $e_{\rm forced}(a)$ in-between the resonances stays  larger than $\tilde{e}$, and our definition of $w$ does not disentangle the two resonances\footnote{Adopting larger $\tilde{e}$ at fixed $e_p$ could modify this behavior. However, it is not clear a priori what value must be assigned to $\tilde{e}$, not least because $e_{\rm forced}(a) \propto e_p$ could stay well above unity in-between the resonances in linear Laplace-Lagrange theory.}. This could be understood, for instance, by looking at the curve for $M_d/m_p = 1$ in Fig. \ref{fig:eforced_a_figure}. These considerations explain why the contours of constant $w/a\inn$ in Fig. \ref{fig:width_map} behave differently for $a\res/a\inn \lesssim 1.5$ compared to $a\res/a\inn \gtrsim 1.5$.

To better understand the behavior of $w/a\inn$, in Appendix \ref{app:res_width_analytics} we derive an analytic expression for the resonance widths showing that, to a good approximation,
\begin{equation}
\frac{w}{a\inn}  
\approx  
\frac{2}{a\inn}  
\left| \frac{B_p(a)/\tilde{e}}{   dA/da } \right|_{a\res} 
\propto 
\frac{e_p}{\tilde{e}} 
\frac{a_p}{a\inn} 
\left( \frac{a\res}{a\inn}   \right)^{-1/2}  , 
\label{eq:width_scaling}
\end{equation}
where the scaling holds for $p=1$ in the limits of $a_p/a\inn \rightarrow 0$ and $a\inn \ll a\res \ll a\out$. First, Equation (\ref{eq:width_scaling}) shows that the width is inversely proportional to the gradient of $A$ at $a\res$. This explains why resonances in proximity of the disk edges are relatively narrow: in the limit of $a\res \rightarrow a\inn, a\out$ we have $A \rightarrow A_d$ which diverges due to edge effects (Fig. \ref{fig:A_a_figure}), and $dA/da$ is very large. Second, we see from Eq. (\ref{eq:width_scaling}) that the width is directly proportional to $B_p \propto e_p$: this makes intuitive sense since $e_p$ controls the amplitude of planetesimal eccentricities (Eq. \ref{eq:eforced}). It follows that more eccentric planets tend to produce wider resonances, provided that $\tilde{e}$ can be chosen independently from $e_p$ (though this is not clear a priori). Third, and more importantly, the scaling of Eq. (\ref{eq:width_scaling}) adequately explains the slopes of the $w/a\inn$ contours: setting $w/a\inn$ to a constant in Eq. (\ref{eq:width_scaling}) yields the scaling $a\res \propto a_p^2$, obvious in Fig. \ref{fig:width_map}. Indeed, by fitting the numerical results in Fig. \ref{fig:width_map} with the functional form of Eq.  (\ref{eq:width_scaling}), we find that the following expression
\begin{equation}
    w \approx 15.3 ~{\rm au} ~  a_{p, 20} ~ a_{\rm res,70}^{-1/2} ~  a_{\rm in, 30}^{1/2} ~ (e_p/\tilde{e})   
\label{eq:num_scaling_width}
\end{equation}
provides an acceptable approximation of the resonance widths for our fiducial disk model (Section \ref{subsec:model_system}).

\section{Example: Application to \HD} 
\label{sec:HD_constraints}

For a given debris disk exhibiting a depletion in its surface density, we can hypothesize that this depletion is due to eccentricity excitation by secular resonances mediated by the gravity of the disk and an unseen planet. We can then employ the characteristics of the secular resonances analyzed in \S \ref{sec:resonances} to constrain the disk-planet parameters that could configure the secular resonances appropriately and produce a depletion similar to the observations. In this section, as an exemplary case, we apply these considerations to the \HD~disk and identify the ``allowed'' parameter space subject to observational constraints. The detailed investigation of the dynamical evolution in models chosen from the allowed parameter space is carried out in the next section.

\subsection{Constraints from gap location}
\label{sec:HD_res_loc}

\begin{figure}[t!]
\epsscale{1.15}
\plotone{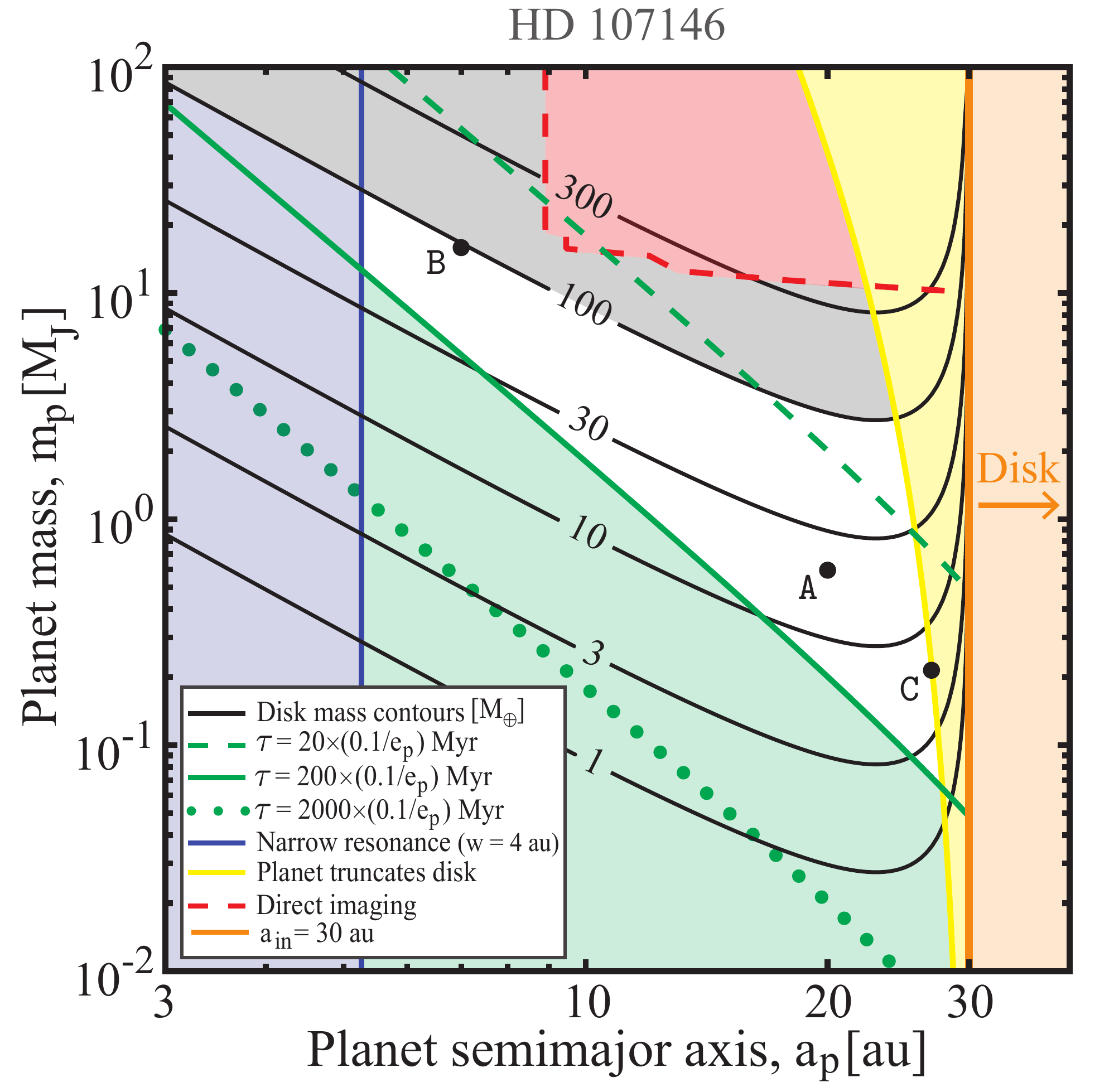}
\caption{Combinations of the planet ($m_p$) and disk masses ($M_d$) as a function of planet semimajor axis $a_p$ that are expected to produce a gap in a \HD-like disk at $70$ au. The curves of constant $M_d$ are shown by the black contours. The grey region is ruled out as the disk would be too massive. The green region shows the excluded region where the eccentricity excitation timescales are much longer than the stellar age. The blue region is ruled out as the resulting resonance width would be much narrower than the observed gap. A planet close to the disk inner edge is ruled out (yellow region) by considerations of overlapping mean motion resonances. The red region is ruled out by direct imaging. The remaining white area represents the region where the disk-planet parameters meet all the above conditions. The lettered points represent the model parameters discussed in Sections \ref{sec:fiducial_results}, \ref{sec:other_configurations} and listed in Table \ref{table:models}. See the text (\S \ref{sec:HD_constraints}) for details.
\label{fig:param_space}}
\end{figure}

As noted in Section \ref{sec:intro}, ALMA observations show that the \HD~disk, spanning from $a\inn \sim30$ au to $a\out \sim150$ au, features a gap centered at $a_g \sim 70-80$ au \citep{ricci2015, marino2018}. Thus, we must choose the disk-planet parameters such that a secular resonance occurs within the depleted region. Here we opt to fix the resonance location at $a\res = 70$ au. The analysis in Section \ref{sec:res_location} then allows us to uniquely determine the ratio $M_d/m_p$ as a function of $a_p/a\inn$, see also Eq. \ref{eq:num_scaling_mass}. In other words, for a given disk mass, we can deduce the planetary mass and semimajor axis that configure the resonance location appropriately (or vice versa). This is displayed by the black solid lines in Figure \ref{fig:param_space} for various values of disk mass (in $M_{\earth}$).

However, the disk mass can not be arbitrarily large and must be constrained. To this end, we note that observations of \HD~have detected around $0.25 M_{\earth}$ of dust at millimeter wavelengths \citep{ricci2015, marino2018}. By extrapolating this up to planetesimals of $\sim 100$ km in diameter the estimated total disk mass is $M_d \sim 100 - 300 M_{\earth}$ (assuming a size distribution with an exponent of $-3.5$, \citet{ricci2015, marino2018}). Here we choose to take $100 M_{\earth}$ as the upper limit of the disk mass. Based on this, we exclude regions in the $(a_p, m_p)$ parameter space that require more massive disks -- see the gray shaded area in the upper part of Figure \ref{fig:param_space}.

\subsection{Constraints from stellar age and disk asymmetry}
\label{subsubsec:age_asymmetry}

We can further constrain the parameter space by considering the age of \HD, which is estimated to be $t_{\rm age} \sim 80 - 200$ Myr \citep{williams2004}. Specifically, we require the timescale for eccentricity excitation at the resonance $\tau$ to be less than around the age of the system, i.e. $\tau \lesssim t_{\rm age}$. From Section \ref{sec:res_timescale}, however, we know that $\tau$ depends not only on the planet's mass and semimajor axis but also on its eccentricity, see Eq. (\ref{eq:axi_timescale}). To this end, we note that ALMA observations have found that the \HD~disk is roughly axisymmetric, with a $2\sigma$ upper limit of $\sim$ 0.03 for the global disk eccentricity \citep{marino2018}. This suggests that the invoked planet must be of relatively low eccentricity. Thus, in what follows, we limit ourselves to $e_p \leq 0.1$.

The green curves  in Figure \ref{fig:param_space} show contours along which the excitation timescale $\tau$ is  $20$, $200$, and $2000$ Myr (dashed, solid, and dotted lines, respectively) at $a\res = 70$ au. The calculations assume $e_p = 0.1$ -- the maximum value of $e_p$ that we consider in our subsequent calculations -- and use the stellar mass of \HD, namely $M_c = 1.09 M_{\odot}$ \citep{watson11}. We first note that by definition $\tau \propto 1/e_p$ (Eq. \ref{eq:axi_timescale}): thus, for less eccentric planets the contours shown in Fig. \ref{fig:param_space} will correspond to longer timescales. Second, recall that $\tau$ is a measure of the time within which  initially circular planetesimal orbits become radial, $e \rightarrow 1$ (\S \ref{sec:res_timescale}). Thus, even if $\tau \gtrsim t_{\rm age}$ for a given planet (such that $e(t_{\rm age}) \lesssim 1$), we might still expect sufficient eccentricity excitation for depletion to be apparent at the resonance within the stellar lifetime. Given these considerations and the uncertainty on the age of the system, we exclude the region in $(a_p, m_p)$ parameter space corresponding to $\tau > 200 \left(0.1/e_p \right)$ Myr. This is illustrated by the green shaded region in Figure \ref{fig:param_space}.

\begin{table*}[th!]
\begin{center}
\caption{Parameters of the disk-planet systems considered in Section \ref{sec:results_new}.
\label{table:models}}
\begin{tabular}{cccccccc}
\tableline
\tableline
Model & $M_d [M_{\earth}]$  & $m_p [M_J]$     & $a_p [au]$ &  $M_d/m_p$ & $e_p$ & $\varpi_p(0)$ & $\tau_{\rm sec} [Myr]$
\\
\tableline
\texttt{A} &  20  & 0.6 & 20 &  $1.05 \times 10^{-1}$ & 0.05  & 0   &  33
\\
\Alowerep &  ... & ... & ... &  ... & 0.025  & ...   & ...
\\
\Ahigherep &  ... & ... & ....  & ... &  0.1  & ...  & ...
\\
\texttt{B} &  95 & 15.8 & 7  & $1.89 \times 10^{-2}$ &  0.05  & ...  & 56
\\ 
\texttt{C} &  6 & 0.2 & 26.93 & $9.44 \times 10^{-2}$ &  ...  & ... & 26
\\
\tableline
\end{tabular}
\tablecomments{The combinations of $M_d$, $m_p$, and $a_p$ (columns 2--4) chosen from the allowed region in Fig. \ref{fig:param_space}. Column 5 presents the disk-to-planet mass ratio. Columns 6--7 present the planet's eccentricity and initial apsidal angle, whose precession period is given in column 8. (a) \texttt{Model A} is the fiducial configuration adopted in this work. (b) Each of the considered models have $\tau \approx 135 \times (0.05/e_p)$ Myr.}
\end{center}
\end{table*}

\subsection{Constraints from gap width}
\label{subsec:constraint_width}

As noted in Section \ref{sec:intro}, the  gap width in the \HD~disk is estimated to be $w_{\rm obs} \approx 40$ au \citep{marino2018}. Given this, the planet's semimajor axis could, in principle, be constrained by using the analysis of resonance widths $w$ in Section \ref{sec:res_width} (recall that $w \propto a_p$, Eq. \ref{eq:num_scaling_width}). However, we recall that the resonance widths as defined in Section \ref{sec:res_width} do not necessarily correspond to the physical width of gaps that we expect to form. Nevertheless, we could still use the definition of $w$ to rule out the range of planetary semimajor axes for which the resonance widths would be negligible, i.e $w/w_{\rm obs} \ll 1$. Here we consider resonance widths to be negligible if $w/w_{\rm obs} \leq 0.1$ (this choice is somewhat arbitrary). The blue solid line in Fig. \ref{fig:param_space} corresponds to $w/w_{\rm obs} = 0.1$; planetary semimajor axes to the left of this line are ruled out (blue shaded region).

\subsection{Considerations of mean-motion resonances}
\label{subsubsec:MMRain}

Finally, we note that the planet can not be arbitrarily close to the disk. This is because the planetary orbit is surrounded by an annular `chaotic zone' wherein particles will be quickly ejected from the system due to overlapping first-order mean motion resonances (MMR). Moreover, the secular approximation of Section \ref{sec:theory} would break down within this zone. The half-width of the chaotic zone on either side of the planetary orbit depends on the planet's mass  \citep{wisdom80, duncan89} such that, to lowest order\footnote{Strictly speaking, Eq. (\ref{eq:mmroverlap}) is valid for circular orbits in the absence of collisions. The chaotic zone is known to broaden with both increasing eccentricity \citep{mustill2012} and due to collisional effects \citep{erika_gap}. For simplicity, we have ignored these effects.}:
\begin{equation}
    \delta a_p \approx 1.3 \left( \frac{m_p}{M_c  + m_p} \right)^{2/7} a_p .
    \label{eq:mmroverlap}
\end{equation}
We thereby can rule out the region in the $(a_p, m_p)$ parameter space wherein the planet's chaotic zone would lie within the disk, i.e. $a_p + \delta a_p > a\inn$. This is illustrated by the yellow shaded region near the right boundary of Fig. \ref{fig:param_space}. Planetary parameters lying along the yellow solid line correspond to $a_p + \delta a_p = a\inn$; thus, they could be responsible for setting the inner disk edge \citep[e.g.][]{quillen_edges} at $a\inn = 30$ au (orange line).

\vskip 1em

We have now identified the `allowed' range of disk-planet parameters that can produce an \HD-like disk structure. This is represented by the white (unshaded) region in Fig. \ref{fig:param_space}, and roughly defined by $a_p$ in the range $\sim 5 - 27$ au, $m_p$ between $\sim 0.1$ and $25 M_J$,  and $3 \lesssim M_d/M_{\earth} \lesssim 100$. Note that the allowed combinations of $m_p$ and $a_p$ are consistent with the limits placed by direct imaging of \HD ~ \citep{apai2008}, see the dashed red curve in Fig. \ref{fig:param_space}. For reference, the combinations of $m_p$, $a_p$ and $M_d$ which we consider later in this work are labelled as \texttt{models} \texttt{A}--\texttt{C} in Fig. \ref{fig:param_space}, see also Table \ref{table:models}. Note that each of these configurations correspond to $\tau \approx 135 \times (0.05/e_p) \rm{Myr}$, and model \texttt{A} represents the fiducial configuration considered next in Section \ref{sec:fiducial_results}.

We remark that in the above discussion we have implicitly ignored the occurrence of an \textit{inner} secular resonance at $\simeq a\inn$; apart from the one already fixed at $a\res = 70$ au in Fig. \ref{fig:param_space}, see \S \ref{sec:res_location}. This can be justified on the grounds that the inner resonance is of very narrow width except if the two resonances are close to each other, which is not the case here (\S\ref{sec:res_width}). As a result, and as we will see next, the inner resonance is irrelevant and does not have any observable effect.

Finally, we point out that equations (\ref{eq:num_scaling_mass}), (\ref{eq:axi_timescale}) and (\ref{eq:num_scaling_width}), combined with Eq. (\ref{eq:mmroverlap}), can be applied to generate an approximate version of Figure \ref{fig:param_space} for any other observed debris disk with a gap.

\section{Evolution of the disk morphology}
\label{sec:results_new}

In the previous section, we identified the combinations of the `allowed' disk-planet parameters that could reproduce the observed depletion in the \HD~disk, see Fig. \ref{fig:param_space}. We now investigate the dynamical evolution of disk-planet systems using some of these parameters. Our specific aims here are two-fold: to illustrate how secular resonances sculpt depleted regions, and to analyze more fully the disk and gap morphology in the course of secular evolution.

\subsection{A Fiducial Configuration}
\label{sec:fiducial_results}

\begin{figure*}[ht!]
\epsscale{1.15}
\plotone{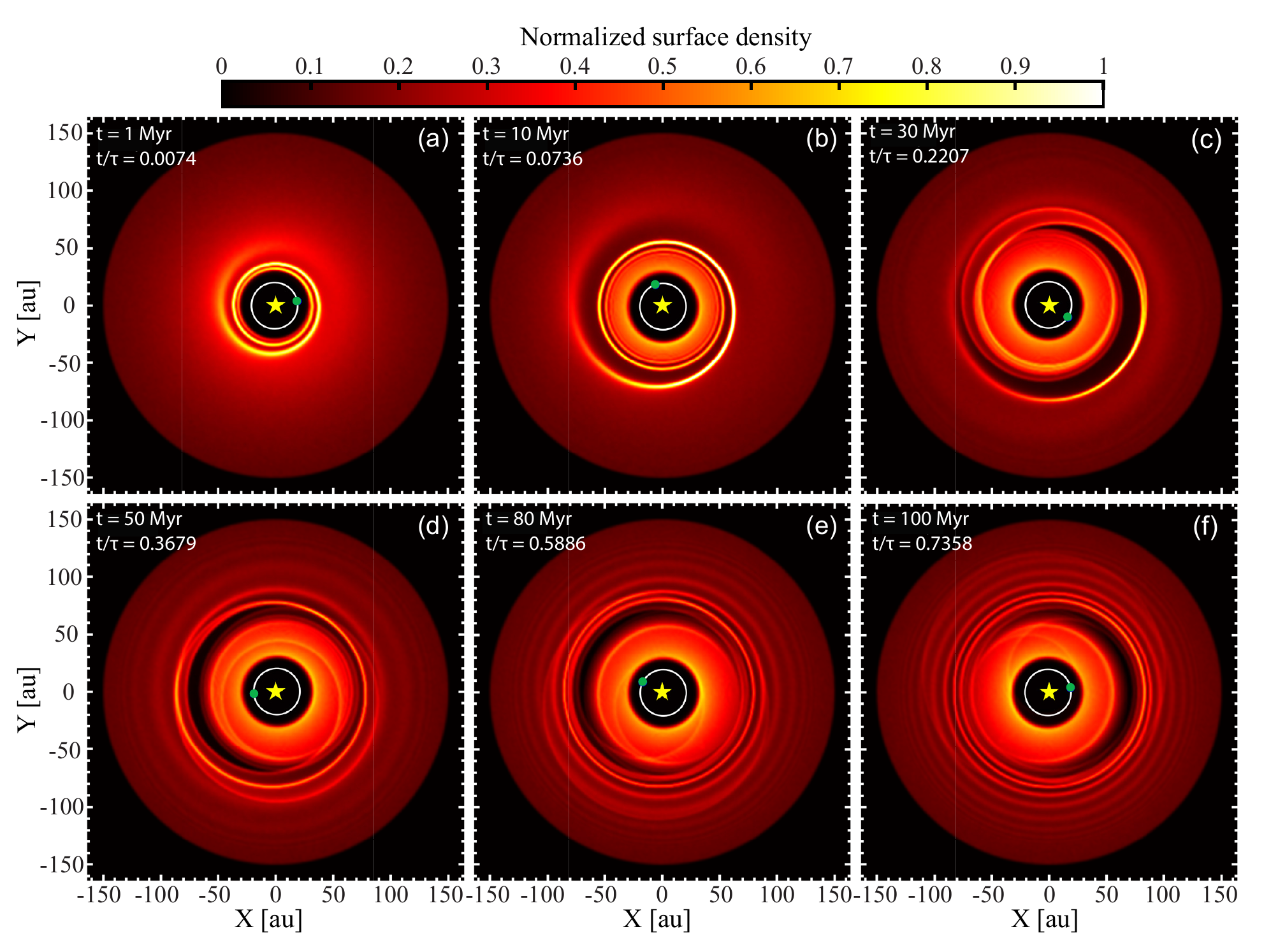}
\caption{Series of two-dimensional snapshots showing the evolution of the (normalized) disk surface density $\Sigma$ in the fiducial model (\texttt{Model A}, Table \ref{table:models}), as derived from the analytically computed dynamical state of planetesimals shown in Fig. \ref{fig:simA_e_w_a_time}. The snapshots correspond to the same moments of time $t$ as in Fig. \ref{fig:simA_e_w_a_time}, and are indicated in each panel for reference.
The time is also indicated relative to $\tau \approx 135$ Myr,  Eq. (\ref{eq:axi_timescale}).
All panels have $400 \times 400$ pixels and share the same surface density scale (and normalization constant) as shown in the colour bar. In each panel the stellar position is marked by the yellow star, while the planet's orbit and its  pericenter position are shown by the white solid line and green circle, respectively. To enhance the resolution of the images, the orbit of each planetesimal ($N = 5000$ in number) has been populated with  $10^{4}$ particles with the same orbital elements but with randomly distributed mean anomalies (see Appendix \ref{app:construct_map}). At early times (panels a, b), the planet launches a trailing spiral wave at the inner disk edge $a\inn$ which is quickly wrapped around the star. By the time the planet has completed around one precession cycle (panel c), a crescent-shaped gap forms around the secular resonance at $a\res = 70$ au, which is both wider and deeper in the direction of planet's pericenter. Beyond this time (panels d--e), the shape of the gap practically remains the same as it precesses while maintaining its coherence with the planet's pericenter. Note that the disk part interior to the gap is offset relative to the exterior part, where a wound spiral pattern is visible at late times (panels d--e). It is also clear that no gap forms around the secular resonance at $\sim a_{\rm in}$. See the text (Section \ref{sec:fiducial_results}) for more details. This figure is available as an animation in the electronic edition of the journal. The animation runs from $t=0$ to $t = \tau \approx 135$ Myr with a duration of $34$ seconds.
\label{fig:DiskMap_simulationA_fiducial} }
\end{figure*}

\begin{figure*}[ht!]
\epsscale{1.15}
\plotone{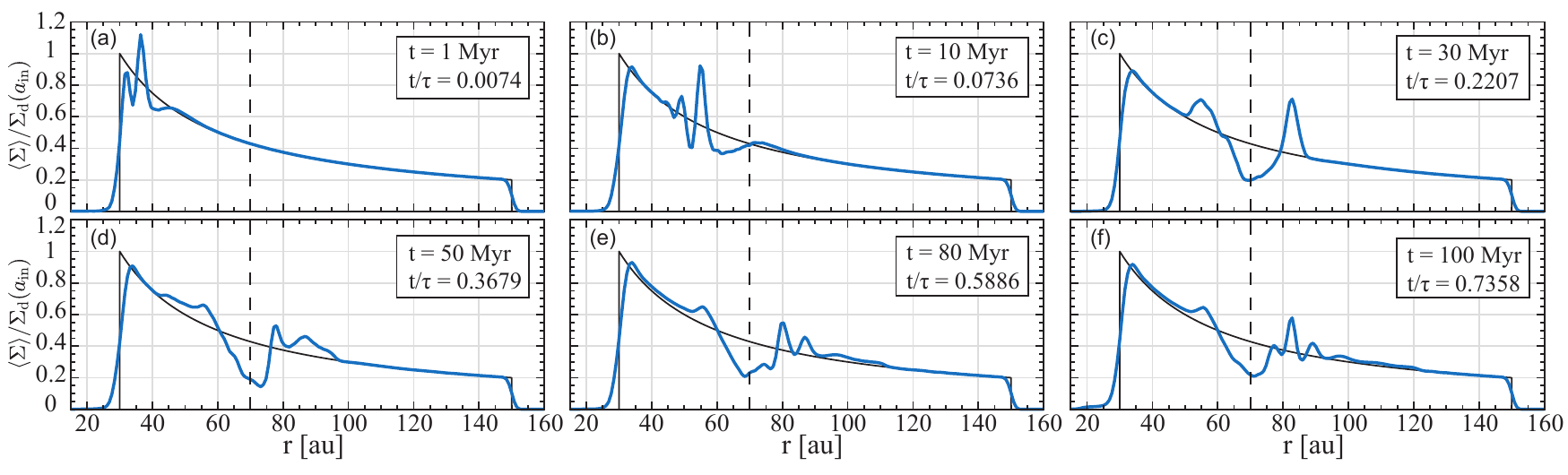}
\caption{The azimuthally-averaged surface density of the disk $\langle \Sigma \rangle$ as a function of radial distance $r$ from the star (solid blue lines). Each panel corresponds to each of the snapshots of the fiducial configuration (Model \texttt{A}, Table \ref{table:models}) shown in Fig. \ref{fig:DiskMap_simulationA_fiducial}. 
The time $t$ of each snapshot is marked in each panel, which is also shown relative to $\tau \approx 135$ Myr for reference.
The results are obtained by splitting the disk into $200$ annular bins (Appendix \ref{app:construct_map}), and are all normalized with respect to the initial analytic surface density $\Sigma_d(a)$ (Eq. \ref{eq:Sigma_d} with $p=1$) at the inner disk edge,  $a = a\inn$. For reference, the normalized profile of the initial $\Sigma_d(a)$ is shown in each panel with the solid black lines. At early times (panels a, b), the overall shape of $\langle \Sigma \rangle$ is similar to the initial profile, but with some peak features around $\sim 40$ au at $1$ Myr and $\sim 60$ au at $10$ Myr, respectively. At all times after $30$ Myr (panels c--e), a clear depletion in the surface density is evident around the location of the secular resonance ($a\res = 70$ au, dashed vertical lines). One can see that the width and the depth of the depletion are effectively constant in time (panels c--e). Note also the peak structure in the density just exterior to the depletion in panels (c)--(e). See the text (Section \ref{sec:fiducial_results}) for more details. This figure is available as an animation in the electronic edition of the journal. The animation runs from $t=0$ to $t = \tau \approx 135$ Myr with a duration of $34$ seconds.
\label{fig:AvDensity_simulationA_fiducial} }
\end{figure*}

We begin by presenting results showing the evolution of the disk surface density in the fiducial configuration, i.e. model \texttt{A} (see Table \ref{table:models}). We recall that model \texttt{A} is the configuration that was considered in Section \ref{subsec:resonance}, where we discussed the temporal evolution of planetesimal eccentricities and apsidal angles as a function of semimajor axis -- see Fig. \ref{fig:simA_e_w_a_time}. To this end, we convert the orbital element distributions of planetesimals shown in Fig. \ref{fig:simA_e_w_a_time} -- which, we remind, were determined \textit{analytically} using equations (\ref{eq:e_t_solution}) and (\ref{eq:w_t_solution}) -- into surface density distributions. Technical details about this procedure can be found in Appendix \ref{app:construct_map}, and may be skipped by the reader at first reading.
However, to avoid confusion, we remark that the results presented here (and in subsequent sections) are obtained by the analytical model described in Section \ref{sec:theory} and not by direct $N$-body simulations, which is beyond the scope of this paper.

The resulting maps of the (normalized) disk surface density $\Sigma$ at times corresponding to those in Fig. \ref{fig:simA_e_w_a_time} are shown in Figure \ref{fig:DiskMap_simulationA_fiducial}.
For reference, in this figure we also show the planet's orbit and its pericenter position, which precesses with a period of $\tau_{\rm sec} \equiv 2\pi/A_{d,p} \approx 33$ Myr (Eq. \ref{eq:Adp}). To facilitate the interpretation of our results, in Figure \ref{fig:AvDensity_simulationA_fiducial} we also show the profiles of the azimuthally-averaged disk surface density $\langle \Sigma \rangle$ as a function of radial distance $r$ at the same times as in Fig. \ref{fig:DiskMap_simulationA_fiducial}. Below we provide a detailed description of the different evolutionary stages that we identified.

\textbf{Stage 1 ($0 \leq t \lesssim \tau_{\rm sec}$):} 
At early times, the disk quickly evolves away from its initial axisymmetric state by developing a trailing spiral structure (see Figs. \ref{fig:DiskMap_simulationA_fiducial}a, b). This spiral structure initially starts off at the inner disk edge and propagates radially outwards with time as it wraps around the star; see also the animated version of Fig. \ref{fig:DiskMap_simulationA_fiducial}. For instance, by $1$ Myr at least two windings are noticeable (Fig. \ref{fig:DiskMap_simulationA_fiducial}a), with the outermost prominent spiral arm occurring at $\sim 40$ au. This arm moves out to $\sim 60$ au by $10$ Myr (Fig. \ref{fig:DiskMap_simulationA_fiducial}b). A complementary view of this behavior is provided by Figs. \ref{fig:AvDensity_simulationA_fiducial}(a),(b).

We note that the outermost portion of the spiral is associated with planetesimal orbits that have attained their maximum eccentricity, i.e. have completed half a precession period -- see Fig. \ref{fig:simA_e_w_a_time}. Interior to this, the spirals become difficult to discern since planetesimals in this region have completed more than one precession period and their orbits are phase-mixed, i.e.  $\Delta\varpi(a)$ spans the range $[-\pi/2, \pi/2]$ -- see Figs. \ref{fig:simA_e_w_a_time}(a), (b). As a result, the surface density distribution interior to the outermost spiral looks roughly axisymmetric; see e.g. panel (b) of Fig. \ref{fig:DiskMap_simulationA_fiducial}. We also note that the spiral propagates outwards at a slower rate as it extends to larger radii; see panels (a)--(c) of Fig. \ref{fig:DiskMap_simulationA_fiducial} and its animated version. This follows from the fact that the planetesimal precession rate is a decreasing function of the semimajor axis (Fig. \ref{fig:A_a_figure}).

We remark that the behavior described thus far shows some parallels with the findings of \citet{wyattSPIRAL05},  which showed that an eccentric planet launches a spiral wave which propagates throughout a \textit{massless} disk. The main difference is that, in our setup, the spiral wave extends out to only about a radius of $70$ au and not to the outer disk edge (as would happen in a massless disk), see Fig. \ref{fig:DiskMap_simulationA_fiducial}. This is to be expected, since in our model planetesimal dynamics is dominated by the planet only within $\approx 70$ au, beyond which the disk gravity becomes important -- see Fig. \ref{fig:A_a_figure} and \S \ref{subsec:resonance}.

\textbf{Stage 2 ($t \sim \tau_{\rm sec}$):} 
By the time the planet has nearly completed its first precession cycle, the disk develops a clear depletion in its surface density, which effectively splits the disk into an internal and an external part (Figs. \ref{fig:DiskMap_simulationA_fiducial}c, \ref{fig:AvDensity_simulationA_fiducial}c). The depletion occurs around the location of the secular resonance, i.e. at $a\res = 70$ au, where the system was designed to emplace one -- see \S \ref{sec:HD_constraints}. The appearance of the gap is evidently correlated with the excitation of planetesimal eccentricities at and around $a\res$, where $e = t/\tau \approx 0.22$ by $30$ Myr (Fig. \ref{fig:simA_e_w_a_time}c).

An interesting feature of the gap is that it is of a crescent shape which points in the direction of the planet's pericenter (Fig. \ref{fig:DiskMap_simulationA_fiducial}c). In other words, the gap is asymmetric in the azimuthal direction such that it is wider and deeper towards the planetary pericenter. This asymmetry is associated with the inner and outer disk components being offset relative to the star in opposite directions (Fig. \ref{fig:DiskMap_simulationA_fiducial}c). Indeed, the inner part forms an eccentric structure which is apsidally aligned with the planet while the outer part is anti-aligned (see also Section \ref{subsec:resonance}) -- the latter though is difficult to discern in Fig. \ref{fig:DiskMap_simulationA_fiducial} due to the smaller eccentricities in the outer parts (Fig. \ref{fig:simA_e_w_a_time}). Nevertheless, by simply looking at the azimuthally-averaged density profile we find that the gap has a radial width of $\sim 20$ au (measured relative to the initial density profile, Fig. \ref{fig:AvDensity_simulationA_fiducial}c). Looking at Fig. \ref{fig:AvDensity_simulationA_fiducial}(c), it is also clear that this region is not depleted fully but only partially -- by about a factor of two relative to the initial density distribution.

Finally, we note that the gap is surrounded by narrow overdense regions,  with the one just exterior to the gap being sharper than that interior to it (see Figs.  \ref{fig:DiskMap_simulationA_fiducial}c, \ref{fig:AvDensity_simulationA_fiducial}c). These overdensities correspond to the apocentric positions of planetesimals with semimajor axes in the depleted region. The contrast between the sharpness of the overdensities is mainly due to the apsidal angles of planetesimals at $a \lesssim a\res$ being more phase-mixed than at $a \gtrsim a\res$ (Fig. \ref{fig:simA_e_w_a_time}c). This also justifies why these sharp overdensities are transients: they taper with time as planesimal orbits around the resonance are perturbed further (see panels d--e in Figs. \ref{fig:simA_e_w_a_time}, \ref{fig:DiskMap_simulationA_fiducial}, and  \ref{fig:AvDensity_simulationA_fiducial}).

\textbf{Stage 3 ($\tau_{\rm sec} \lesssim t \lesssim \tau$):}
Further into the evolution, the structure of the gap practically remains invariant without being significantly affected by the continued growth of eccentricity around $a\res = 70$ au (see panels d--e in Figs. \ref{fig:simA_e_w_a_time}, \ref{fig:DiskMap_simulationA_fiducial}). Indeed, the gap maintains its crescent shape along with its alignment with the planet as it co-precesses with the planet's apsidal line.

At the same time, since the inner component of the disk precesses much faster than the outer component (Fig. \ref{fig:A_a_figure}), the degree of offset between them varies as the system evolves. This causes the gap width $w_g$ to fluctuate in time, see e.g. Figs. \ref{fig:AvDensity_simulationA_fiducial}(d)--(e), with a time-averaged value of $w_g \approx 18.13 \pm 1.04 $ au. Looking at Figs. \ref{fig:AvDensity_simulationA_fiducial}(d)--(e), it is also clear that the gap depth remains roughly constant such that, in a time-averaged sense, about $50 \pm  3 \%$ of the initial density is depleted at the resonance.

Note that, at this stage, i.e. at $t \gtrsim \tau_{\rm sec}$, at least one secular period has elapsed for planetesimals interior to the depletion, causing them to settle into a lopsided, precessing coherent structure (Figs. \ref{fig:DiskMap_simulationA_fiducial}d--e). It is also noticeable that this structure reveals little or no evidence for surface density asymmetry between its apocenter and pericenter directions, as would have otherwise been the case if the disk were \textit{massless} \citep[i.e. pericenter or apocenter glow; see][]{wyattetal99, wyattSPIRAL05, margaretapocenter16}. This can be understood by noting that in this region, although planetesimal dynamics is dominated by the planet, the disk gravity renders the forced eccentricity to be more of a constant with semimajor axis rather than scaling as $1/a$ (see Figs. \ref{fig:A_a_figure}, \ref{fig:simA_e_w_a_time}). This hinders the occurrence of a pericenter or apocenter glow (for a more detailed discussion, see section 2.4 in \citet{wyattSPIRAL05}).

On the other hand, planetesimal orbits exterior to the depletion have not yet had the time to be randomly populated in phase (Fig. \ref{fig:simA_e_w_a_time}). Hence, a spiral pattern develops in this region as planetesimals undergo eccentricity oscillations. The spirals appear to wrap almost entirely around the star, and these are more noticeable closer to the depletion than to the outer disk edge (Figs. \ref{fig:DiskMap_simulationA_fiducial}d--e). This can also be seen in Figs. \ref{fig:AvDensity_simulationA_fiducial}(d)--(e) as a series of narrow peaks in the radial profile of $\langle \Sigma \rangle$. This behavior can be understood by noting that planetesimals closer to the outer disk edge have smaller eccentricities (e.g. Fig. \ref{fig:simA_e_w_a_time}) and that their orbits are quickly phase-mixed as a result of their rapid orbital precession due to disk edge effects, particularly at $a \gtrsim 130$ au (e.g. Fig. \ref{fig:A_a_figure}, \S \ref{sec:dist_function}). Relatedly, if we were to evolve the system for longer, planetesimals exterior to the depletion would become phase-mixed and the spiral structure would fade away. We note that, depending on the resolution of observations, the spirals in this region may or may not be visible.

\vskip 1em

Before moving on, we note that already by $1$ Myr into the evolution, planetesimal eccentricities around the inner secular resonance (i.e. $a\res \approx a\inn$) are excited to $\approx 1$; see e.g. Fig. \ref{fig:simA_e_w_a_time}(a). Evidently, however, this occurs over a narrow radial range that it does not lead to the emergence of a gap (see Figs. \ref{fig:DiskMap_simulationA_fiducial},  \ref{fig:AvDensity_simulationA_fiducial}), in agreement with our expectations from Section \ref{sec:res_width}. This also justifies our assertion in Section \ref{sec:HD_constraints} about ignoring the occurrence of an inner secular resonance for the purposes of Fig. \ref{fig:param_space}.

\subsection{Parameter variation} 
\label{sec:variations_HD107} 

We now analyze the variation of the disk morphology associated with varying the disk-planet parameters relative to the fiducial values (\texttt{Model A}).


\subsubsection{Variation of the planetary semimajor axis $a_p$}
\label{sec:other_configurations}

\begin{figure}[ht!]
\epsscale{1.20}
\plotone{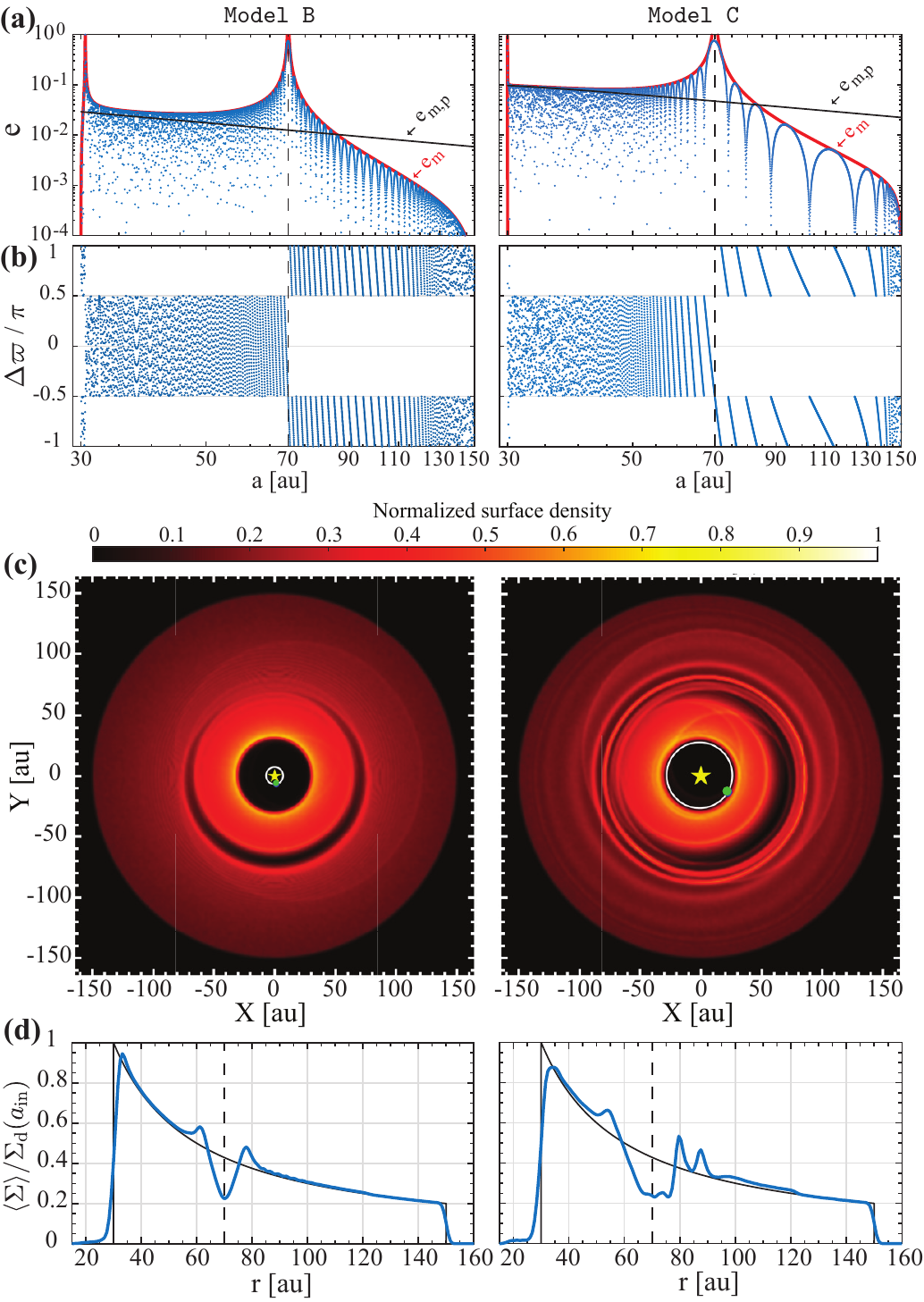}
\caption{Summary of results for \texttt{Model B} (small $a_p$, left column) and \texttt{Model C} (large $a_p$, right column), see Table \ref{table:models}. The results are shown after $100$ Myr of evolution, corresponding to $t/\tau \approx 0.74$ for both models. Rows (a) and (b) show the planetesimal eccentricities and apsidal angles (relative to that of the planet) as a function of semimajor axis, respectively, which are determined analytically using Eqs. (\ref{eq:e_t_solution}), (\ref{eq:w_t_solution}). The corresponding snapshots of the disk surface density and radial profiles of the azimuthally-averaged surface density are shown in the rows (c) and (d), respectively -- see Appendix \ref{app:construct_map} for details. All other notations are the same as in Figs. \ref{fig:simA_e_w_a_time}, \ref{fig:DiskMap_simulationA_fiducial} and \ref{fig:AvDensity_simulationA_fiducial}. One can see that wider gaps are carved around the secular resonance at $a\res = 70$ au when the planet is closer to the disk inner edge than to the star. It is also evident that the resultant gaps are asymmetric and of approximately the same depth in both models. See the text (\S \ref{sec:other_configurations}) for more details. \label{fig:Compilation_simB_to_E} }
\end{figure}

We first consider the effects of varying the planetary semimajor axis $a_p$ which, we remind, all else being kept the same, is equivalent to changing the ratio $M_d/m_p$ (\S \ref{sec:res_location}, \S \ref{sec:HD_constraints}). For ease of comparison, we choose the combinations of $a_p$, $m_p$ and $M_d$ from Fig. \ref{fig:param_space} such that they yield the same eccentricity excitation timescale at the secular resonance $\tau$ as in model \texttt{A}. The parameters of the chosen models, which we label as \texttt{B} and \texttt{C}, are listed in Table \ref{table:models} and are marked on Fig. \ref{fig:param_space}. Note that the planet in \texttt{Model C} could be responsible for truncating the disk at $a\inn = 30$ au; see \S \ref{subsubsec:MMRain}.

Generally, we find that the evolution of the disk morphology in each of models \texttt{B} and \texttt{C} proceeds in a similar manner as in the fiducial model (i.e. stages 1--3 in \S \ref{sec:fiducial_results}). Indeed, we observe the same qualitative behaviour: the launching of a spiral arm at $a\inn$ and its outward propagation in time, the sculpting of a crescent-shaped gap around $a\res = 70$ au by $\sim \tau_{\rm sec}$, the development of a spiral pattern exterior to the depletion at $t \gtrsim \tau_{\rm sec}$ and its subsequent potential disappearance at late times (depending on the period of secular precession at $a \gtrsim a\res$).

Figure \ref{fig:Compilation_simB_to_E} summarizes the snapshots of models \texttt{B} and \texttt{C} at $100$ Myr (i.e. $t/\tau \approx 0.74$) into their evolution. A comparison of the results shown in this figure with those of \texttt{Model A} (Figs. \ref{fig:DiskMap_simulationA_fiducial}f, \ref{fig:AvDensity_simulationA_fiducial}f) indicate that the only obvious difference is in terms of the radial width of the gaps $w_g$. Indeed, the gap is radially narrower when the planet is closer to the star than to the inner disk edge: for $a_p = 7$ au (i.e. \texttt{Model B}), on time-average, $w_g \approx  11.32 \pm  0.05$ au, while for $a_p = 26.93$ au (i.e. \texttt{Model C}) we have $w_g \approx 20 \pm  2$ au. This dependence will be investigated in the future (Rafikov \& Sefilian, in preparation), though for now we note that it is in qualitative agreement with our expectation from Section \ref{sec:res_width} regarding the resonance widths. Finally, we note that the gap depth is not affected by variations in planetary semimajor axis: on average, about a half of the initial density is depleted around the secular resonance regardless of $a_p$.

\subsubsection{Variation of the planetary eccentricity $e_p$}
\label{sec:effect_of_ep_variation}

The models presented thus far assumed the same planetary eccentricity of $e_p = 0.05$. To examine its effect on the disk morphology, we considered the evolution in otherwise identical setups but differing in the value of $e_p$ by a factor of two from model \texttt{A}. These are referred to as models \Alowerep~(with $e_p = 0.025$) and \Ahigherep~(with $e_p = 0.1$) in Table \ref{table:models}.

Once again, we found that the evolution of the disk morphology qualitatively follows the same stages outlined in Section \ref{sec:fiducial_results}, but on a shorter timescale when the planet is more eccentric (recall that $\tau \propto 1/e_p$, Eq. \ref{eq:axi_timescale}). Additionally, we identified subtle differences in the structure of the spiral arms with increasing $e_p$. First, the spiral initially launched at $a\inn$ by the planet became more open for larger $e_p$ -- in agreement with the results of \citet{wyattSPIRAL05}. Second, and relatedly, the spirals beyond the gap became more prominent with increasing $e_p$ due to the higher forced eccentricities in that region.

\begin{figure}[t!]
\epsscale{1.20}
\plotone{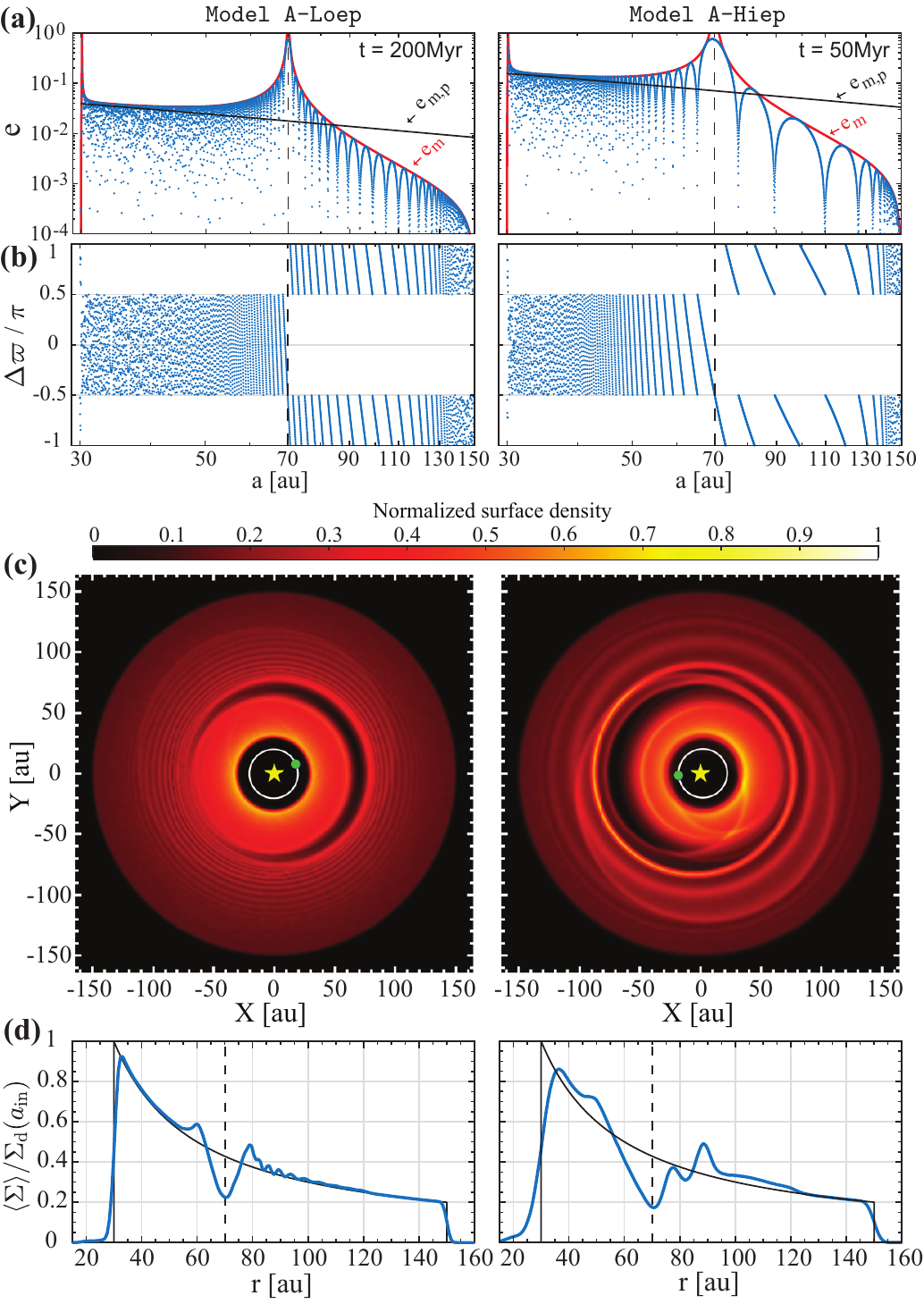}
\caption{Similar to Fig. \ref{fig:Compilation_simB_to_E}, but for models \Alowerep~(left panels) and \Ahigherep~(right panels); see Table \ref{table:models}. Models \Alowerep~and \Ahigherep~are identical to the fiducial model \texttt{A}, except that they are initiated with planets with eccentricities that are lower and higher by a factor of two than in model \texttt{A} (i.e. $e_p$ of $0.025$ and $0.10$), respectively. For ease of comparison, results for each model are shown at different times (as indicated in the top panels) such that they both correspond to $t/\tau \approx 0.74$. One can see that increasing $e_p$ leads to a wider gap around the secular resonance at $a\res = 70$ au, without significantly affecting the asymmetric shape of the gap and its depth. See the text (\S \ref{sec:effect_of_ep_variation}) for more details.
\label{fig:SimA_with_lowerhigher_eplanet} }
\end{figure}

More importantly, however, we found that more eccentric planets give rise to wider gaps\footnote{We defer a quantitative characterization of this dependence to future work (Rafikov \& Sefilian, in prep.)} -- in qualitative agreement with our expectations from Section \ref{sec:res_width}, see Eq. (\ref{eq:width_scaling}). Indeed, on time-average, we find that $w_g \approx 12.8 \pm 0.2$ au when $e_p = 0.025$, and $w_g \approx 24.6 \pm 2.8$ au when $e_p = 0.10$. This can be seen in Figure \ref{fig:SimA_with_lowerhigher_eplanet}, where we summarize the results for models \Alowerep~and \Ahigherep. Note that, for ease of comparison, the results are shown at different times such that $t/\tau(e_p) \approx 0.74$ for both models -- the results must be compared with those of model \texttt{A} at $100$ Myr (Figs. \ref{fig:DiskMap_simulationA_fiducial}, \ref{fig:AvDensity_simulationA_fiducial}). Looking at Fig. \ref{fig:SimA_with_lowerhigher_eplanet}, it is also evident that variations in $e_p$ do not significantly affect the fractional depth of the gap. Note also that, while planets with lower $e_p$ reduce the offset of the inner disk component, the gap retains its non-axisymmetric feature. This is largely related to the fact that for narrower gaps a smaller offset suffices for the inner component to occupy about the same fraction of the gap.

\subsubsection{Variations with disk and planet masses}
\label{sec:effect_of_mdmp_variation}

We now discuss the effects of varying the disk and planet masses while keeping other parameters unchanged. To begin with, we first recall that this requires varying both $M_d$ and $m_p$ simultaneously, i.e. while keeping $M_d/m_p$ constant, to ensure that the secular resonance location where a gap is expected to form remains the same (i.e. $a\res = 70$ au); see \S \ref{sec:res_location} and \S \ref{sec:HD_res_loc}. In Figure \ref{fig:param_space}, this is equivalent to moving vertically up or down relative to any of the simulation setups we have considered thus far.

As we know from Section \ref{sec:theory}, the secular precession rates scale linearly with masses (Eqs. \ref{eq:Aplanet} -- \ref{eq:Adp}),  whereas the forced eccentricities depend only on the ratio $M_d/m_p$ (Eq. \ref{eq:eforced}). Thus, varying the disk and planet masses (while $M_d/m_p = \rm{cte}$) should only change the secular evolution timescale, but not the details of the secular dynamics. This simply is a restatement of the fact that scaling both $M_d$ and $m_p$ does not affect the relative strength of perturbations due to the disk and the planet. Consequently, if we increase both the disk and planet masses in any of our simulations, then the very same dynamical end-states -- hence, disk morphology -- will be achieved within shorter timescales, and vice versa. We note that, in principle, this scaling rule applies as long as $M_d, m_p \ll M_c$, since otherwise the Laplace-Lagrange description in Section \ref{sec:theory} becomes unreliable \citep{mur99}. However, looking at Figure \ref{fig:param_space} we see that this limitation is not a concern in our case: the most massive allowed planet has $m_p \sim 10^{-2} M_c$.

\subsubsection{Variations with the mass distribution in the disk}
\label{sec:effect_of_p}

Our calculations so far have assumed a disk with density profile $\Sigma_d \propto 1/a$, i.e. with a power-law index of $p=1$ in Eq. (\ref{eq:Sigma_d}). We now discuss how our results would change for different values of $p$, when all else is kept the same. Since the slope of the surface density $p$ effectively controls the precession rate of both the planetesimals and the planet  (Eqs. \ref{eq:Adisc}, \ref{eq:Adp}), it is natural to expect that the location of the secular resonance will shift as the mass distribution in the disk is varied; see also Eq. (\ref{eq:resonance_condition_literal}). We found that this is indeed the case, and we further confirmed that it does not qualitatively affect the evolutionary stages presented in Section \ref{sec:fiducial_results}.

We generally find that when $a\inn \ll a\res \ll a\out$, the resonance location shifts at most by only about 10 per cent as $p$ is varied between $0.5$ and $1.5$. However, the direction in which the resonance shifts in a given setup is rather subtle to characterize for the following reasons. First, larger values of $p$ lead to larger $A_{d,p}$ (and \textit{vice versa}) as now more mass will be concentrated in the inner disk parts than in the outer regions, causing the planet to precess at a faster rate. Second -- and relatedly --  the disk induced precession rate of planetesimals $A_d$ at $a \gg a\inn$ decreases in absolute magnitude, since it is proportional to the local surface density of the disk (Eq. \ref{eq:Adisc})\footnote{We recall that $A_d(a)$ depends also on $p$ through the coefficient $\psi_1$; however, the latter changes by less than a factor of 2 within the range $0.5 \leq p \leq 1.5$ \citep[e.g.][]{sil15}.}. To summarize, varying $p$ has opposite effects on $A_{d,p}$ and $|A_d|$, and it is the detailed balance between these two effects that determines whether the resonance shifts outwards or inwards in a given setup, see Eq. (\ref{eq:resonance_condition}). For the parameters of \HD~in Figure \ref{fig:param_space}, we find that the resonance shifts inwards from its nominal location, i.e. $a\res = 70$ au, when a larger value for $p$ is adopted  (and vice versa). Thus if we were to generate a version of Figure \ref{fig:param_space} with e.g. $p=1.5$ rather than $p=1$, the values of $M_d$ required to reinstate the resonance at $a\res= 70$ au would be a factor of $\sim 1.1 $ lower.

\section{Discussion}   \label{sec:discussion}

The results of previous sections show that the secular interaction between a low-eccentricity planet and an external, co-planar debris disk can lead to the formation of a gap in the disk. This occurs through the excitation of planetesimal eccentricities at around one of the two secular resonances arising due to the combined gravitational influence of the disk\footnote{Recall that in this paper we ignore the non-axisymmetric component of the disk gravity. See Section \ref{subsubsec:non_axi_effects} for further discussion of this point.} and the planet. The novelty of this mechanism is that it requires the presence of only a single planet interior to a less massive disk, and is also robust, in the sense that it operates over a wide range of parameters.

As an example, we applied our model to the \HD~disk and investigated the general features of the disk and gap morphology in the course of secular evolution. In the following, we first discuss (in a general context) how the results of our model compare with the observed features in \HD~(\S \ref{disc:comparison_HD107146}). We also discuss the application of our model to other systems (\S \ref{disc:other_disks}). Finally, we discuss the implications of our study for determining the masses of debris disks (\S \ref{disc:disk_mass_determination}), and for their dynamical modeling in general (\S \ref{disc:diskg_imp}),

\subsection{Comparison with observed structure in \HD}
\label{disc:comparison_HD107146}

By applying our model to \HD, we have shown that a gap can be readily sculpted at the observed location, i.e. around $70$ au \citep{marino2018}, for a wide range of planet-disk parameters; see e.g. Fig. \ref{fig:param_space}, Section \ref{sec:results_new}. Additionally, our results show that the produced gaps invariably have a fractional depth of about $0.5$ (Section \ref{sec:results_new}), which is consistent with that observed in \HD~\citep{marino2018}. While these results are encouraging, there are some issues with our model that need to be highlighted when it comes to comparing with the observational data of \HD~\citep{marino2018}.

First, as already mentioned in Section \ref{subsubsec:age_asymmetry}, ALMA observations of \HD~indicate that its disk is axisymmetric and characterized by a circular gap \citep{marino2018}. Our model, however, produces gaps that are asymmetric in the azimuthal direction (Section \ref{sec:results_new}), with the disk surface density being depleted to a greater extent and over a wider region in the direction of planet's pericenter. We further found that the gap asymmetry can not be mitigated, as one might naively expect, by adopting lower values for the planetary eccentricity -- see Section \ref{sec:effect_of_ep_variation}.

Second, as already stated in Section \ref{subsec:constraint_width}, the observed gap in \HD~is $\sim 40$ au wide. This is larger by about a factor of two compared to the gap in our fiducial configuration (Section \ref{sec:fiducial_results}). In principle, our model can yield such wide gaps with a combination of high-eccentricity and large semimajor axis for the planetary orbit; see Sections \ref{sec:other_configurations} and \ref{sec:effect_of_ep_variation}. However, this would also impose more notable  non-axisymmetric structure on the disk which, given the discussion above, is problematic for \HD. Thus the conclusion is that, within the limitations of our model (for a detailed discussion, see Section \ref{sec:limitations_future}), it is difficult to sculpt a gap as wide and as axisymmetric as that in \HD~without invoking additional processes. We discuss a way in which a wider gap could form as a result of disk mass depletion and secular resonance sweeping in Section \ref{subsec:collision_sweeping}.

Third, observations of \HD~indicate that the surface brightness of the outer and inner rings are comparable \citep[see fig. 2 in][]{marino2018}. Since sub-mm dust emission at a distance $r$ scales as $T(r)\propto r^{-1/2}$ (assuming black body emission in the Rayleigh-Jeans limit), this observation suggests an \textit{increasing} surface density with radius, which may seem unnatural in the context of protoplanetary disks. As a result, this has been taken as evidence for collisional depletion of planetesimals in the inner disk regions \citep{ricci2015, yelverton2018}.  Thus, if our collisionless model were applied to any physically realistic profile (i.e. with $p>0$, Eq. \ref{eq:Sigma_d}), it is unlikely that we would reproduce the observed brightness peaks. However, it is possible that a shallower density slope than $p=1$ could generate comparable brightness peaks at times $t \sim \tau_{\rm sec}$, when our model produces an overdensity just exterior to the depletion (see Stage 2 in  \S \ref{sec:fiducial_results}).

The above discussion suggests that although our mechanism acting alone can produce a structure qualitatively similar to that observed in \HD, it does not provide a quantitative interpretation of the observations. However, we re-emphasize that our aim in this work was not to provide a complete description of the \HD~disk, but rather to provide a  proof-of-concept for our mechanism and its feasibility. We also stress that the limitations of our simple model need to be assessed before making any definitive conclusions (see \S \ref{sec:limitations_future} for a detailed discussion). Our results serve as a starting point to guide future, more comprehensive studies which aim to match the observations of the \HD~disk, or any other disk with an observed gap.

Given the potential ubiquity of gaps in debris disks \citep[e.g.][]{kennedywyatt14, marino2020}, it is also possible that future surveys will reveal a sample of disks with asymmetric gaps. Two potential candidates for such systems are \HDN~\citep{marino2019} and \HDnew~\citep{marino2020}, which we discuss next.

\subsection{Application to other systems}
\label{disc:other_disks}

\subsubsection{\HDN}
\label{sec:HD92945discussion}

We first consider the system \HDN~\citep{golimowski11} which is often viewed as a sibling to \HD~in many ways. Both systems not only have stars with similar masses and ages ($1 M_{\odot}$ and $100-300$ Myr, \citet{plavchan09}), but also their disks show some similarities in terms of their radial structure. Indeed, ALMA observations of \citet{marino2019} show that the \HDN~disk, extending from $\sim 50$ to $140$ au, is double-peaked with a gap centered at about $\sim 73$ au, roughly coincident with that in \HD. However, and in contrast to \HD, the gap in \HDN~appears to be asymmetric and is relatively narrow with an estimated width of $20^{+10}_{-8}$ au \citep{marino2019}.

\begin{figure}[t!]
\epsscale{1.15}
\plotone{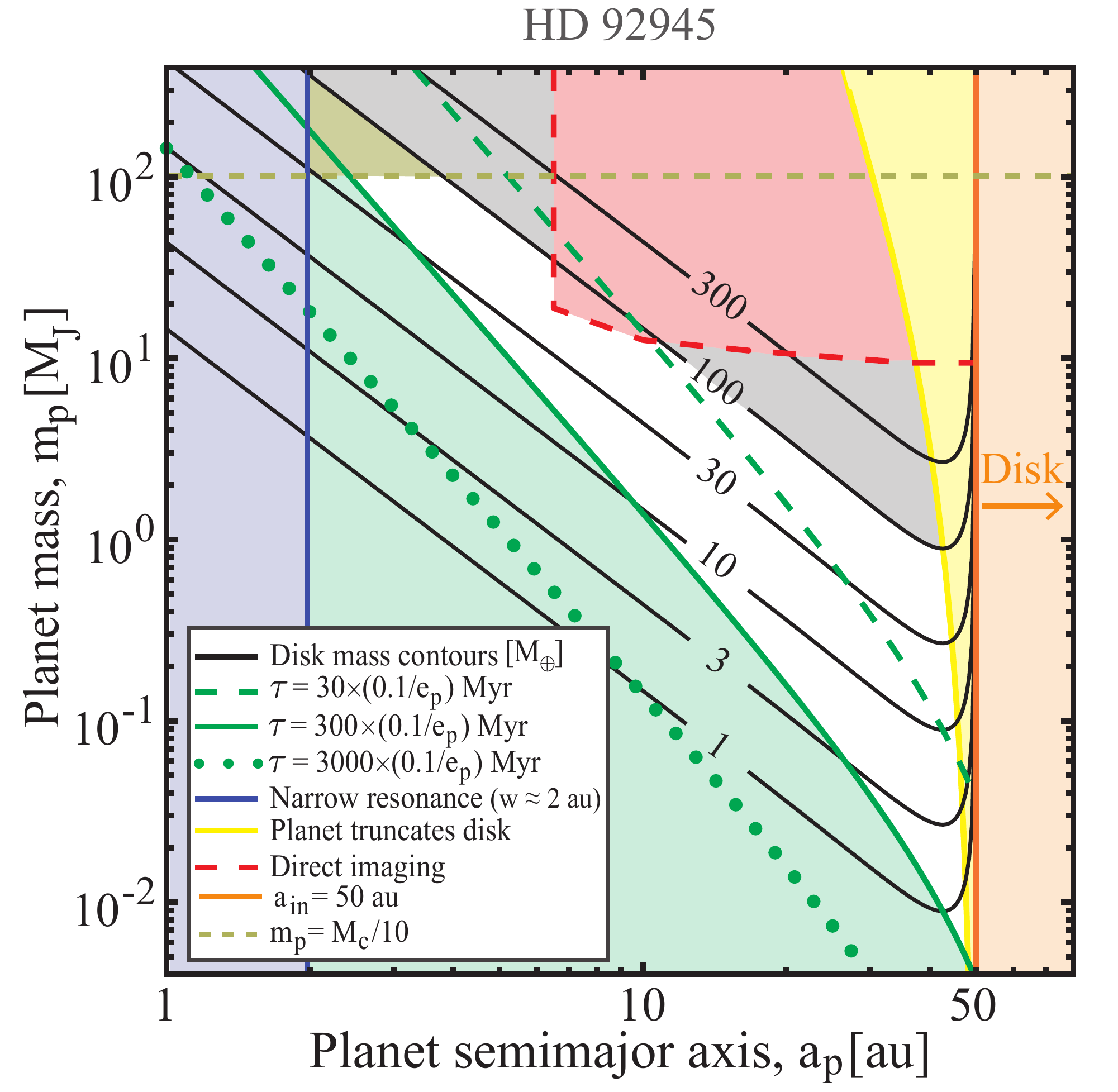}
\caption{Similar to Fig. \ref{fig:param_space}, but for a \HDN-like disk. The white region represents the disk-planet parameters that place a secular resonance at $73$ au such that it acts on a time-scale less than the stellar age (i.e. $\leq 300$ Myr) and is wide enough to have an observable effect. All other notations and exclusion criteria are similar to those in Fig. \ref{fig:param_space}, except that here we have also excluded planet masses that exceed one tenth of the central star mass, i.e. $m_p \geq M_c/10$ (olive shaded region in the top part of parameter space). See the text (\S \ref{sec:HD92945discussion}) for details.
\label{fig:param_spaceHD92945}}
\end{figure}

These features speak in favor of our model, so we could use our results (\S \ref{sec:resonances}) to determine the properties of the planet and disk such that the gap is sculpted by secular resonances. Figure \ref{fig:param_spaceHD92945} summarizes the results of our analysis (following a similar reasoning as for \HD~in Section \ref{sec:HD_constraints}). We find that a companion with a semimajor axis $a_p$ in the range $\sim 3 - 50$ au and mass $m_p$ between $\sim 10^{-2}$ and $10^2 M_J$ can produce a wide enough gap  at the observed location within the stellar age, provided that $1 \lesssim  M_d/M_{\earth} \lesssim 100$ -- see the white region in Fig. \ref{fig:param_spaceHD92945}. These limits are in agreement with (i) direct imaging constraints \citep[][red curve in Fig. \ref{fig:param_spaceHD92945}]{biller13}, and (ii) disk mass estimates of $\sim 100 - 200 M_{\earth}$ derived from collisional models \citep{marino2019}.

Finally, we note that since the inner disk edge in \HDN~is located at $\sim 50$ au, i.e. further out than in \HD, it is possible for the planet to be on a more distant orbit than in \HD~(Fig. \ref{fig:param_spaceHD92945}). However, we confirmed that this is only necessary if the true gap width is towards the upper end of its estimated range (recall that increasing $a_p/a\inn$ in our model leads to wider gaps). For instance, we find that invoking a planet similar to that in \texttt{Model A} (but with a disk of mass $M_d \approx 16.4 M_{\earth}$) produces a $\sim 16$ au wide gap, which is comparable to that observed. Future observations of this system could help to put better constraints on the disk mass and planetary properties.

\subsubsection{\HDnew}
\label{sec:HDnew_MD}

We next consider \HDnew, a $50 - 700$ Myr old F5V star, which hosts a debris disk \citep{marino2020} as well as one brown dwarf companion, \HDnew~B, detected using direct imaging \citep{milli17}. ALMA observations of \citet{marino2020} show that this disk, extending from $\sim 30$ to $180$ au, features an asymmetric $\sim 27$ au wide gap centered at $\sim 75$ au. Given that \HDnew~B orbits interior to the disk with $a_p \sim 11$ au \citep{delorme17}, this system is ideally suited to test whether our model can reproduce the observed gap.

To assess this, we adopt the minimum possible mass of \HDnew~B \citep[$\sim 12 M_J$,][]{delorme17} and calculate, using Eq. (\ref{eq:resonance_condition}), the disk mass that would place a secular resonance at the observed gap location, i.e. $a\res = 75$ au. Assuming a surface density profile with $p=1$ (Eq. \ref{eq:Sigma_d}), we find that the required disk mass is $M_d \approx 170 M_\oplus$; see also Eq. (\ref{eq:num_scaling_mass}). This is roughly consistent with the disk mass estimates of \citet{marino2020} based on collisional models. Moreover, we also confirmed that the gap width $w_g$ obtained from our model agrees well with that observed: adopting the best-fitting eccentricity of \HDnew~B, $e_p \sim 0.15$ \citep{marino2020}, we find that $w_g \approx 26$ au after $\sim 20$ Myr of evolution. If future observations with better resolution confirm that the gap in the \HDnew~disk is indeed wider towards the companion's pericenter position, this will then provide a strong support to our model.

Finally, we note that recent analyses of \HDnew~have indicated that it is likely that this system harbors a second inner companion at $\sim 2$ au \citep{grandjean19, marino2020}. While in this work we only considered single-planet systems, our model may easily be extended to two-planet systems (or more).
In this case, depending on the strength of perturbations from the companion(s), our results both in general (e.g. Section \ref{sec:theory}) and for \HDnew~may or may not be affected significantly.
Although such an analysis is beyond our scope here, we briefly discuss this caveat in Section \ref{sec:multiplanetsystems}.

\subsection{Implications for disk mass estimates}
\label{disc:disk_mass_determination}

Our results may be used to infer the presence of a yet-undetected planet in any system harboring a double-ringed debris disk. The inferences are, of course, degenerate with the assumed system parameters but, more importantly, they are subject to the condition that there be sufficient mass in the disk (Sections \ref{sec:resonances}, \ref{sec:HD_constraints}). Thus, the detection of planets with the inferred properties will not only provide strong support to our model, but also -- and more importantly -- provide a unique way to indirectly measure the total mass of the debris disk $M_d$ (see e.g. Section \ref{sec:HDnew_MD}). This is particularly appealing, considering the fact that $M_d$ can not be accessed using other techniques -- not least without invoking theoretical collisional models to extrapolate observed dust masses to the unobservable larger planetesimals that carry most of the disk mass \citep[see][for a detailed discussion]{krivovwyatt20}. This represents a promising avenue to consider in the future, in particular with the advent of new generation instruments such as JWST which could detect planets with $m_p \lesssim 10 M_J$ at $a_p \sim 10$ au separations. Conversely, the results of Section \ref{sec:resonances} may be used to investigate whether or not the debris disk of a known planet-hosting system should have a gap. Future observations of such systems e.g. with ALMA looking for evidence -- or lack thereof -- of a gap could help in constraining  the total disk mass.

\subsection{The importance of disk self-gravity in dynamical modelling of debris disks} 
\label{disc:diskg_imp}

The study presented here has further consequences beyond an explanation of gap formation in debris disks. Particularly, our findings strongly emphasize the need to account for the (self-)gravitational effects of disks in studies of planet-debris disk interaction. As we showed in this study, the end-state of secular interactions between a single planet and a disk having only a modest amount of mass can be radically different from the naive expectations based on a massless disk. Indeed, if it were not for the disk gravity in our model, secular resonances would have not been established and so no gap would have formed in the disk -- at least not without invoking two or more planets \citep[e.g. as done by][]{yelverton2018}, or a single but precessing planet \citep{pearcewyatt15}.

This also highlights an important caveat related to the dynamical modelling of debris disks in general. While studies treating debris disks as a collection of massless particles seem to successfully reproduce a large variety of observed disk features by invoking unseen planets \citep[e.g. see reviews by][]{krivovreview2010, wyatt18review}, their inferences about the underlying planetary system architecture  may be compromised. The inclusion of disk gravity would -- at least -- impose modifications on the masses and orbital properties, if not numbers, of invoked planets. Thus, caution must be exercised in the interpretation of observed disk structures when the disk mass is ignored.

Recently, \citet{dong19} raised a similar point when it comes to ascribing observed morphologies of disks (assumed to be massless) to single planets in situations where the potential presence of a second planet is ignored. We urge a similar analysis to be performed by considering a natural hypothesis of having non-zero disk mass in contrast to the potential presence of additional planets. Although this is beyond the scope of our current work, the formalism outlined in Section \ref{sec:theory} could provide a useful starting point for such an analysis. To summarize, the inclusion of disk self-gravity in studies of planet-disk interactions should be considered in dynamical modelling of debris disks.

\section{Limitations and future work} \label{sec:limitations_future}

We now review some of our model assumptions and limitations, and discuss how relaxing them would affect our results. We plan to address these issues in future papers of this series.

\subsection{Disk model assumptions}
\label{subsec:effect_of_nonaxi}

\subsubsection{Treating planetesimals as test-particles}
\label{subsubsec:simplified_sims}

In this work we treated planetesimals as massless test-particles, and analyzed their secular evolution under the influence of gravity from both the planet and the debris disk. To this end, we modelled the debris disk as being \textit{passive}: that is, as a rigid slab that provides \textit{fixed} axisymmetric gravitational potential (see Eq. \ref{eq:RdRp} and \S \ref{sec:theory}, disk non-axisymmetry is discussed next in \S \ref{subsubsec:non_axi_effects}). Thus, at first glance, it appears that instead of the planetesimals to be contributing to the collective potential of the disk, they are enslaved by the fixed disk potential given in Eq. (\ref{eq:RdRp}). In reality, though, these two approaches are subtly similar. This is because the orbit-averaged disturbing function for a planetesimal of mass $m_j$ due to all other $N$ massive planetesimals in a disk -- in the continuum limit (i.e. $N\rightarrow\infty, m_j \sim N^{-1}$) -- is equivalent to that in Eq. (\ref{eq:RdRp}). This can be verified by somewhat tedious but straightforward calculation which requires softening the gravitational interaction between massive planetesimals, integrating radially over all  planetesimals, and taking the limit of zero softening \citep{hahn2003, SR19}.

To further justify this equivalence, we simulated the secular dynamics of disk-planet systems by modelling the disk as a swarm of $N$ massive planetesimals, each represented as a ring\footnote{Recall that orbit-averaging is equivalent to smearing particles into massive rings along their orbits, where the line-density of each ring is inversely proportional to the orbital velocity of each particle \citep{mur99}.}, that interact via softened gravity \citep[e.g.][]{hahn2003, JTgauss, bat12}. We found that simulations carried out with negligible softening parameter accurately reproduce the analytical solutions presented in Section \ref{subsec:evol_eq_theory} (which is, of course, possible only when the non-axisymmetric perturbations due to simulated disk particles are neglected, i.e. as in \S \ref{sec:theory}). We will present further details about this softened `$N$-ring' method in an upcoming work (Paper II).

\subsubsection{Non-axisymmetric component of disk gravity}
\label{subsubsec:non_axi_effects}

A major limitation of this work is that we only accounted for the axisymmetric contribution of the disk gravity, ignoring its non-axisymmetric component (Section \ref{sec:theory}). That is to say, our model does not account for the non-axisymmetric perturbations that disk particles can exert both among themselves and onto the planet (see \S \ref{subsubsec:simplified_sims}), even though we find that the disk naturally develops non-axisymmetry (Sections \ref{sec:fiducial_results}, \ref{sec:variations_HD107}). This omission allowed us to elucidate the key effects of disk gravity (semi-)analytically. This comes at the expense of reduced coupling within the system that inhibits the exchange of angular momentum between the disk and planet. Thus, the outlined theory serves as a first step towards a comprehensive understanding of the role played by disk gravity and its observational implications.

Previous studies of gravitating disk-planet systems (which include the full gravitational effects of disk particles) have shown that an eccentric planet could launch a long, one-armed, spiral density wave at a secular resonance in the disk \citep{wardhahn1998, hahn2003, hahn2008}. Such spiral waves propagate away from the resonance location as trailing waves with pattern speed equal to the planetary precession rate. These waves also transfer angular momentum from the disk to the planet in a way that damps the planet's eccentricity, without affecting its semimajor axis\footnote{This process is referred to in the literature as ``resonant friction'' \citep{tre98} or ``secular resonant damping" \citep{wardhahnprotostars}.} \citep{goldreich1980, wardhahn1998, tre98, wardhahnprotostars}.

Our idealized model is not designed to capture the full richness of such dynamical phenomena. Thus, a more sophisticated analysis is crucial, and will be the subject of future work (Paper II, in preparation). For now, we note that the non-axisymmetric component of disk gravity is not going to qualitatively affect the gap-forming picture. This is because the divergence of eccentricities at the resonance ensues from the commensurability between planetesimal and planetary precession rates, while the torques due to the planet's and disk's non-axisymmetric potentials are non-zero \citep[e.g.][]{sil15, silsbeekepler, irina18}. Nevertheless, the generation of long spiral waves exterior to the depleted region may affect the disk structure and its evolution; this could be of observational relevance. Additionally, the damping of planetary eccentricity could reduce the gap asymmetry observed in our simulations via lowering $e_{\rm forced}$ over time, especially in the inner disk parts. Preliminary simulations carried out with the softened `$N$-ring' model confirm these expectations (Paper II).

\subsection{Collisional depletion of planetesimals}
\label{subsec:collision_sweeping}

We modelled the debris disk as an ensemble of collisionless planetesimals. In practice, once the disk is sufficiently stirred, planetesimals collide and break up into smaller fragments, initiating a collisional cascade \citep[e.g.][]{wyatt08collisionsreview}. In this process, colliding planetesimals are gradually ground to dust until they are removed from the system by radiation effects; causing the disk mass to collisionally deplete over time.

We expect collisions to preferentially deplete the disk density around the secular resonance (where  $e \rightarrow 1$ and relative velocities between planetesimals are high), in addition to the purely dynamical depletion illustrated in Sections \ref{sec:fiducial_results}, \ref{sec:variations_HD107}. This may enhance the gap depths arising from our collisionless model. Collisional evolution may also contribute to widening the gaps resulting from our model. This can be understood as follows: as the total disk mass is depleted over time, the system's precession frequencies get altered, modifying the location of the secular resonances in a time-dependent way\footnote{We note that this could also happen if the planet migrates, either inwards or outwards, due to some physical process not considered here.} \citep[e.g.][]{hep80, ward81, nagasawa00}. Looking at Fig. \ref{fig:resonance_map}, we can infer that the resonance would sweep through the disk outwards as $M_d$ decreases, potentially producing a wider gap than in our model (as then eccentricities could be excited over a larger range in semimajor axis).
This could be important e.g. for the \HD~disk, for which our fiducial model produces gaps that are narrower than observed (see \S \ref{disc:comparison_HD107146}). Furthermore, we expect the shape of the resulting gap to provide information on the initial and final disk masses along with the history of mass loss. We defer detailed investigation of collisional effects to future work.

\subsection{Coplanarity of the disk-planet system}
\label{subsec:coplanar}

Another assumption of our model is the coplanarity of the debris disk and the planetary orbit, which can be easily relaxed in future studies.  Generally, however, we believe that a small but non-zero relative inclination (e.g. $\lesssim 5^\circ$) between the planet and disk particles would not affect our results for eccentricity dynamics \citep[e.g.][]{pearce14}. This is because the evolution of eccentricities $e$ and inclinations $I$ are decoupled from each other when $e, I \ll 1$ \citep{mur99}. Nevertheless, it is possible for planetesimal inclinations -- similar to eccentricities -- to be excited significantly at \textit{inclination resonances} \citep[e.g.][]{hahn2003, hahnbending}, where the precession rates of both planet's and planetesimal's longitudes of ascending node are commensurate. 
In principle, this could happen when the planet is initially inclined with respect to a razor-thin disk, or when the planet lies in the mid-plane of a puffed-up disk that is populated by planetesimals with non-zero inclination dispersion.
Future studies should investigate this intriguing phenomenon.

\subsection{Secular approximation}
\label{subsec:limitation_LL}

We limited the expansion of the secular disturbing function to second order in eccentricities (\S \ref{sec:theory}). Hence, our results are only approximate at high eccentricities, e.g. in the vicinity of the secular resonances, where it is necessary to include higher-order terms in the disturbing function \citep[e.g. see][]{ST19}. Such an exercise would, primarily, limit the eccentricity amplitude at the resonance \citep{malhotra98}. Nevertheless, it seems unlikely that this would affect the gap formation. For instance, from Figs. \ref{fig:DiskMap_simulationA_fiducial}, \ref{fig:AvDensity_simulationA_fiducial}, we can see that the gap is already well-developed when eccentricities at the resonance are still rather modest, i.e. $ e \sim 0.2$. Higher-order terms, however, could give rise to mild quantitative differences in terms of the dynamical timescales, e.g. period of eccentricity oscillations.

We also ignored mean-motion resonances between the planet and the planetesimals. Previously, \citet{maryam} found in simulations of synthetic debris disks that gaps can be carved at the 2:1 MMR with an internal low-$e$ planet \citep[$e_p\lesssim 0.1$, see also][]{regaly18}. In our simulated systems, this can occur around $\simeq a\inn$. However, as the authors explain, MMR gaps will be blurred or even washed out by high-eccentricity planetesimal orbits further out in the disk. In our case, this could be easily achieved by planetesimals in the vicinity of the secular resonance.

\subsection{Extension to multi-planet systems} 
\label{sec:multiplanetsystems}

Finally, we only considered what is arguably the least complex planetary system architecture: a single planet orbiting interior to a massive disk. However, the model presented in Section \ref{sec:theory} may easily be extended to systems of two (or more) planets interior to the disk. The presence of additional planet(s) may or may not affect our results, depending on the perturbation strength of the additional planet(s).

In a two-planet system, for instance, it is straightforward to expect that our results would remain roughly the same if the perturbations due to the additional planet are negligible, e.g. if it is much less massive and closer to the central object than its counterpart.
The extreme of course is a system  where the additional companion overshadows the gravitational effects of the disk -- even if the latter is relatively massive, say, with $M_d \sim 100 M_{\earth}$.
Such a case would be reminiscent of the setup in \citet{yelverton2018}, where the authors show that two planets carve a crescent-shaped gap -- similar to that we find in our study (Section \ref{sec:results_new}) -- centered around \textit{one} of the two secular resonances they establish within an external, \textit{massless} disk.
The transition between these two extreme cases remains an interesting scenario to explore. In this case it may be possible to carve either two or a single but broader gap in the disk, depending on the properties of the secular resonances  of the ``two planets + massive disk" system which, in principle, can  feature up to four resonances (where two of them will be near $a\inn$ due to disk edge effects, see \S \ref{sec:SR_at_ain}).
A detailed investigation of the potential effects of an additional planet on our results is beyond our scope here and is best deferred to a future study. Nevertheless, we acknowledge that it could be important for the location (if not number) of secular resonances and thus is crucial for constraining the disk-planet parameters based on imaged gap structures.

\section{Summary}  
\label{sec:summary}

In this work we explored the secular interaction between an eccentric planet and an external self-gravitating debris disk, using a simplified analytic model. The model is simplified in the sense that it only accounts for the axisymmetric component of the disk (self)-gravity, ignoring its non-axisymmetric contribution. Despite this limitation, however, this is the first time (to our knowledge) that the effects of disk gravity have been considered analytically in such detail in the context of debris disks. We used the analytic model to assess the possibility of forming gaps in debris disks through excitation of planetesimal eccentricities by the secular apsidal resonances of the system. We summarize our key results below.
\begin{enumerate}[label=(\roman*)]
    
    \item When the debris disk is less massive than the planet, $10^{-4} \lesssim M_d/m_p \lesssim 1$, the combined gravity of the disk and the planet can mediate the establishment of two secular apsidal resonances in the disk. 
    
    \item We map out the behavior of the characteristics of the secular resonances -- i.e. locations, time-scales, and widths -- as a function of the disk and planet parameters. In particular, we find that one of the secular resonances can lead to the formation of an observable gap over a broad region of parameter space.
    
    \item As an example we applied our results to \HD~and \HDN, and showed how the properties of a yet-undetected planet, together with the mass of the debris disk, can be constrained to produce a gap at the observed location. In the case of \HDnew, we find that the directly imaged companion can sculpt the observed gap if the debris disk is $\approx 170 M_{\earth}$ in mass.
    
    \item By investigating the secular evolution in such systems, we identified three distinct evolutionary stages which occur on timescales measured relative to the planetary precession period. We find that the gap forms by the time the planet has completed around one precesional cycle, on a timescale of tens of Myr.
    
    \item Independent of the system parameters, the gap carved around the secular resonance is asymmetric: it is both wider and deeper in the direction of the planetary pericenter. Additionally, its fractional depth is always about $0.5$. The gap width, however, increases with increasing planetary semimajor axis and/or eccentricity.

    \item More generally, our results suggest that the gravitational potential of debris disks can have a notable effect on the secular evolution of debris particles. We advocate the inclusion of disk gravity in studies of planet-debris disk interactions.

\end{enumerate}

The mechanism presented here represents what is arguably the simplest pathway to forming gaps in debris disks, akin to those observed in \HD, \HDN~and \HDnew. It may indeed obviate the need for invoking more complicated scenarios, e.g. multiple planets interior to or within the disk.

Finally, we remark that the present work should be envisaged as a first step towards an in-depth exploration of the effects of disk gravity in planet-debris disk interactions. In a forthcoming paper (Paper II), we will extend our current calculations using numerical techniques to properly account for the full gravitational effects of the disk. In the future, we also plan to investigate the role of disk gravity in shaping debris disk morphologies other than gaps.

\acknowledgments
We express our gratitude to Jihad Touma and Mher Kazandjian for a number of insightful discussions in the early phases of this work. We are also grateful to Sebastian Marino for useful discussions, and the referee for a positive report and constructive comments on the manuscript.
A.A.S. thanks the Gates Cambridge Trust for support toward his doctoral studies (OPP1144). R.R.R. acknowledges financial support through the NASA grant 15-XRP15-2-0139, STFC grant ST/T00049X/1, and John N. Bahcall Fellowship. This article is made open access thanks to the Bill \& Melinda Gates foundation.

\appendix

\section{Disturbing function of planet due to disk gravity}
\label{app:effect_of_disk}

To calculate the secular disturbing function $R_{d,p}$ of the planet due to an external disk, we use equations (4)--(6) from \citet{SR19} for the case of unsoftened gravity. Strictly speaking, these equations represent the continuum version of the classical Laplace-Lagrange theory \citep[e.g.][]{mur99}, and are valid for arbitrary profiles of disk surface density $\Sigma_d(a)$, eccentricity $e_d(a)$, and apsidal angle $\varpi_d(a)$.

For the purposes of this work, we consider the disk to be apse-aligned (i.e. $d\varpi_d/da = 0$) and have surface density $\Sigma_d(a)$ given by Eq. (\ref{eq:Sigma_d}). For future use in Paper II, we also adopt a power-law scaling for the disk eccentricity given by
\begin{equation}
    e_d(a) = e_0 \left( \frac{a\out}{a} \right)^q  
    \label{eq:ed_PL}
\end{equation}
for $a\inn \leq a_ \leq a\out$. Plugging these ansatzes into Eqs. (4)--(6) of \citet{SR19} it can be shown, after some algebra, that $R_{d,p}$ is given by:
\begin{eqnarray}
    R_{d,p} 
    & = &  n_p a_p^2 \left[ \frac{1}{2} A_{d,p} e_p^2 + B_{d,p} e_p \cos\left(\varpi_p - \varpi_d\right) \right]   ,
\end{eqnarray}
with
\begin{eqnarray}
A_{d,p}(a_p) &=& 2\pi \frac{G \Sigma_d(a\inn) }{n_p a_p} 
\frac{a\inn}{a_p} \phi_1 ,
\label{eq:Adp_appendix}
\\
B_{d,p}(a_p) &=& \pi \frac{G \Sigma_d(a\inn)}{n_p a_p} 
\frac{a\inn}{a_p} e_d(a\inn)  \phi_2 .
\label{eq:Bdp_appendix}
\end{eqnarray}
Here $A_{d,p}$ represents the free precession rate of the planetary orbit in the disk potential, while $B_{d,p}$ represents the torque exerted on the planet by the non-axisymmetric component of the disk gravity (which we have neglected in this work, \S \ref{subsubsec:non_axi_effects}). The effects of the latter will be explored in the future (Paper II).

The coefficients $\phi_1$ and $\phi_2$ appearing in Eqs. (\ref{eq:Adp_appendix}) and  (\ref{eq:Bdp_appendix}), respectively, are given by:
\begin{eqnarray}
\phi_1 &=& \frac{1}{4} \left( \frac{a_p}{a\inn}  \right)^{1-p} 
\int\limits_{a_p/a\out}^{a_p/a\inn} \alpha^{p-1} b_{3/2}^{(1)}(\alpha) d\alpha ,
\nonumber
\\
& &
=  \frac{3}{4} \left( \frac{a_p}{a\inn}  \right)^2 \frac{1-\delta^{-1-p}}{p+1} \phi_1^c , 
\label{eq:phi1_appendix}
\\
\phi_2 &=& -\frac{1}{2} \left( \frac{a_p}{a\inn} \right)^{1-p-q}
\int\limits_{a_p/a\out}^{a_p/a\inn} \alpha^{p+q-1} b_{3/2}^{(2)}(\alpha) d\alpha ,
\nonumber
\\
& & = -\frac{15}{8} \left( \frac{a_p}{a\inn}\right)^3 \frac{1-\delta^{-2-p-q}}{p+q+2}  \phi_2^c , 
\label{eq:phi2_appendix}
\end{eqnarray}
where $\delta \equiv a\out/a\inn$.
\begin{figure}[t!]
\epsscale{1}
\plotone{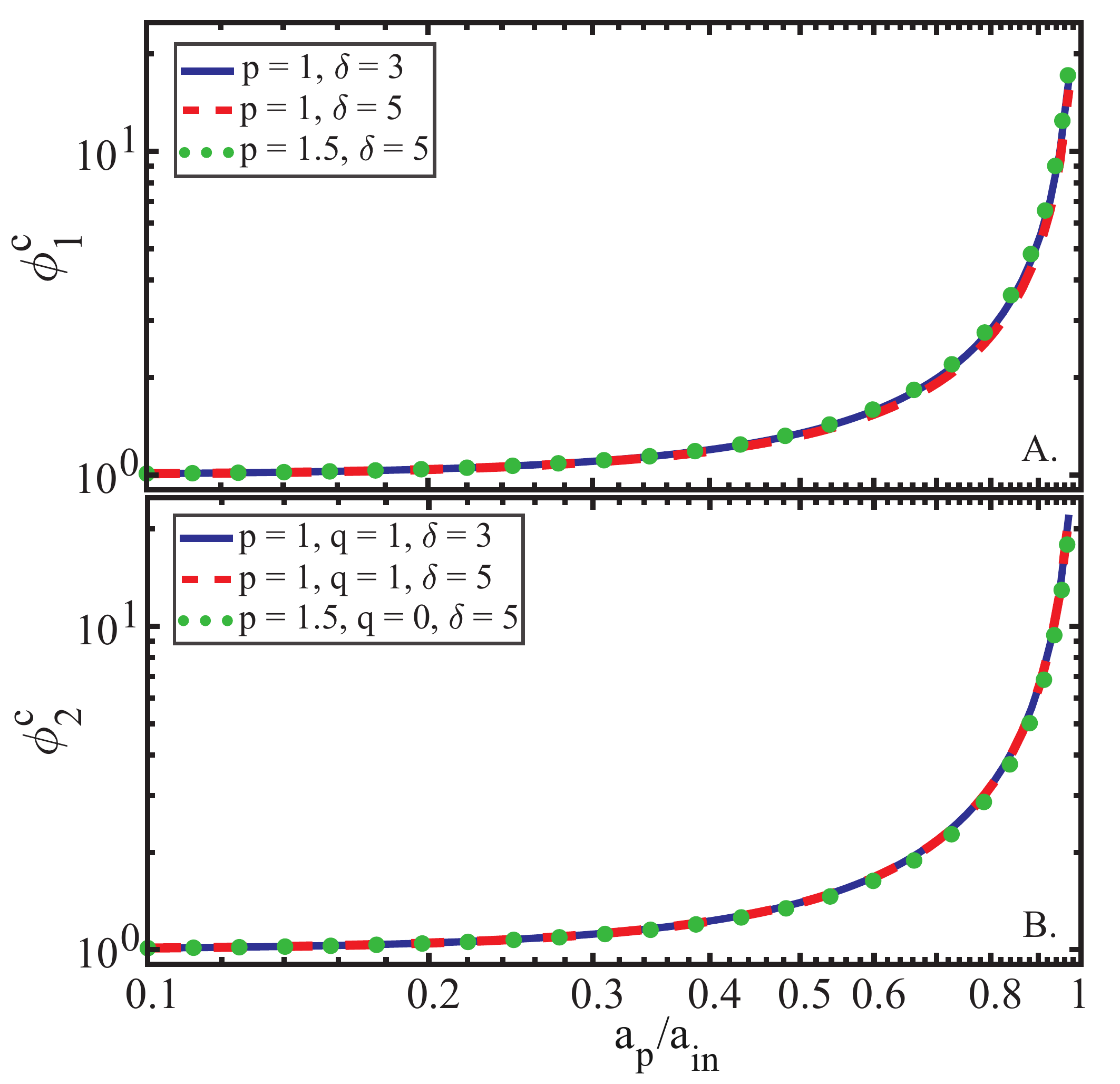}
\caption{The behavior of the correction factors $\phi_1^c$ (panel A, Eq. \ref{eq:phi1c_appendix}) and $\phi_2^c$ (panel B, Eq. \ref{eq:phi2c_appendix}) as a function of $a_p/a\inn$. The calculations assume different disk models specified by the values of $p$, $q$ and $\delta =a\out/a\inn$ as explained in legend. Both $\phi_1^c$ and $\phi_2^c$ approach unity as $a_p/a\inn \rightarrow 0$, and they diverge as $a_p/a\inn \rightarrow 1$.
\label{fig:phi_i_12_app}}
\end{figure}
Here, the second lines in both (\ref{eq:phi1_appendix}) and (\ref{eq:phi2_appendix}) are obtained by performing the integrals appearing in the definitions of $\phi_1$ and $\phi_2$ assuming $\alpha \rightarrow 0$; that is, $b_{3/2}^{(1)}(\alpha) \approx 3\alpha$ and $b_{3/2}^{(2)}(\alpha) \approx (15/4)\alpha^2$. Thus, the coefficients $\phi_1^c$ and $\phi_2^c$ in Eqs. (\ref{eq:phi1_appendix}) and (\ref{eq:phi2_appendix}) represent correction factors accounting for the contribution of disk annuli close to the planet, i.e. higher order terms in $b_{3/2}^{(m)}(\alpha)$. It is straightforward to show that
\begin{eqnarray}
\phi_1^c &=& \frac{1}{3} \frac{p+1}{1-\delta^{-p-1}} \frac{a\inn}{a_p} 
\int\limits_{1}^{\delta} u^{-p-1} b_{3/2}^{(1)}\left( \frac{1}{u} \frac{a_p}{a\inn} \right) du ,
\label{eq:phi1c_appendix}
\\
\phi_2^c &=& \frac{4}{15} \frac{p+q+2}{1-\delta^{-p-q-2}} \left(\frac{a\inn}{a_p} \right)^2 
\int\limits_{1}^{\delta} u^{-p-q-1} b_{3/2}^{(2)}\left( \frac{1}{u} \frac{a_p}{a\inn} \right) du .
\label{eq:phi2c_appendix}
\end{eqnarray}
Figure \ref{fig:phi_i_12_app} shows the behavior of $\phi_1^c$  and $\phi_2^c$ as a function of $a_p/a\inn$, computed for different values of $p$, $q$ and $\delta$. For clarity, we have plotted the curves of $\phi_1^c$ and $\phi_2^c$ in separate panels. We see that $\phi_i^c$ ($i=1,~2$) mainly depend on $a_p/a\inn$, showing weak dependence on the disk model. Indeed, regardless of ($p, q, \delta$), we have $\phi_i^c \rightarrow 1$ for $a_p/a\inn \rightarrow 0$, while in the limit $a_p/a\inn \rightarrow 1$ we see that $\phi_i^c$ diverge. This divergence follows from the fact that $b_{3/2}^{(m)}(\alpha) \rightarrow (1 - \alpha)^{-2}$ when $\alpha \rightarrow 1$.

Finally, we note that inserting Eqs. (\ref{eq:disk_mass}) and (\ref{eq:phi1_appendix}) into Eq. (\ref{eq:Adp_appendix}) results in the expression for $A_{d,p}$ given by Eq. (\ref{eq:Adp}). A similar expression was found by \citet{petrovich19} \citep[see also][]{ward81, RRptype}.

\section{Analytic expression for resonance widths} 
\label{app:res_width_analytics}

The width $w$ of a given resonance at $a = a\res$ can be approximated by using the fact that
\begin{equation}
A\left( a\res + \frac{w}{2} \right) -  A\left( a\res - \frac{w}{2} \right)
\approx 
w  \times \frac{dA}{da} \bigg|_{a\res} . 
\label{eq:w1}
\end{equation}
Additionally, Equation (\ref{eq:width_numeric}) allows us to write 
\begin{equation}
    A\left( a\res \pm w/2 \right) \approx A_{d,p}  \mp  \tilde{e}^{-1} B_p(a\res) \times \mathrm{sgn}\left[ dA/da \right]_{a\res}  ,
    \label{eq:w2}
\end{equation}
where $\mathrm{sgn}(x) = x/|x|$ is the sign function introduced to account for the fact that resonances occurring at $\simeq a\inn$ have $dA/da>0$, while those further away have $dA/da <0$; see Fig. \ref{fig:A_a_figure}. Substituting Eq. (\ref{eq:w2}) into Eq. (\ref{eq:w1}), we thus arrive at
\begin{equation}
\frac{w}{a\inn} \approx \frac{2}{a\inn}  
\left| \frac{B_p(a) \tilde{e}^{-1} }{   dA/da } \right|_{a\res} .
\label{eq:width_approx_app}
\end{equation}
The above expression can be further simplified by considering the approximate forms of $A_p$ and $A_d$ in the limits of $a_p/a\res \rightarrow 0$ and $a\inn \ll a\res \ll a\out$, respectively. In this case, we can approximate the derivative of $A=A_p+A_d$ in the following fashion 
\begin{equation}
\frac{dA_p}{da}\bigg|_{a_p \ll a}  =   \frac{-7}{2a} A_p, 
\quad
\frac{dA_d}{da}\bigg|_{\psi_1 = cte}  =    \frac{1-2p}{2a} A_d ,
\end{equation}
and expression (\ref{eq:width_approx_app}) reduces to
\begin{equation}
    \frac{w}{a\inn}  \approx  4 \frac{a\res}{a\inn}   \left| \frac{B_p(a\res)  \tilde{e}^{-1} }{7 A_p(a\res) + (2p-1) A_d(a\res)}   \right| .
\label{eq:width_approx_app2}
\end{equation}
Inserting the condition for secular resonances, i.e. Eq. (\ref{eq:resonance_condition}) or Eq. (\ref{eq:num_scaling_mass}), into the above expression for $p = 1$, and taking the limits $a_p/a\inn \rightarrow 0$ (so we can use the asymptotic behavior of $b_{s}^{(m)}(\alpha)$) and $a\inn \ll a\res \ll a\out$, we arrive at the scaling relationship given by Eq. (\ref{eq:width_scaling}).

\section{Constructing maps of disk surface density}
\label{app:construct_map}

Here, we provide some technical details about how we convert the eccentricity-apsidal angle distribution of planetesimals into maps of disk surface density.

We first begin by assigning a mass $m_i$ to each considered planetesimal in a given annulus of the disk (which in this work are $N=5000$ in number, \S \ref{subsec:evol_eq_theory}). Given that in our calculations the planetesimals are initiated on circular orbits, the planetesimal masses can be determined from their initial semimajor axis distribution -- which remains constant in the secular approximation. This can be done by using the relationship $dm(a) = 2\pi a \Sigma_d(a) da$ \citep{statler01, irina18} relating the mass distribution per unit semimajor axis to the density distribution (which in our case is given by Eq. \ref{eq:Sigma_d} with $p=1$, \S \ref{subsec:model_system}). The self-consistency of this initial mass assignment to planetesimals -- which are essentially treated as massless particles in our analytical model (see \S \ref{sec:theory}) -- is discussed in Section \ref{subsubsec:simplified_sims}.

At a given time of the evolution, we then populate every planetesimal's orbit with $N_{np} = 10^{4}$ new particles: each with mass $m_i/N_{np}$, orbital elements similar to the parent planetesimal, but with randomly distributed mean anomalies $l$ between $0$ and $2\pi$. This procedure is motivated by the orbit-averaging principle \citep{mur99}. We also note that this procedure effectively increases the number of evolved planetesimals (from $N$ to $N\times N_{np}$), enhancing the quality of the resultant maps of disk surface density. Next, we numerically solve for each new particle's eccentric anomaly $\epsilon$ using Kepler's equation \citep{mur99},
\begin{equation}
   l = \epsilon - e \sin \epsilon , 
\end{equation} 
and compute the position of each particle along its orbit via \citep{st99, BT}:
\begin{equation}
\begin{pmatrix} 
X  \\  Y 
\end{pmatrix}
= a
\begin{pmatrix} 
\cos \varpi &  - \sin \varpi \\
\sin \varpi &    \cos \varpi 
\end{pmatrix} 
\cdot
\begin{pmatrix} 
\cos \epsilon - e \\ \sqrt{1-e^2} \sin \epsilon
\end{pmatrix}
.
\label{eq:XYcoordinate}
\end{equation}
Finally, we bin the positions of all $N \times N_{sp}$ particles in the Cartesian system centered at the host star (with a resolution of $400\times400$ pixels in this work), compute the total mass per bin and divide by its area to arrive at the disk surface density distribution, $\Sigma$, at a given time. Note that this also allows us to trivially obtain the azimuthally-averaged surface density profile $\langle \Sigma \rangle$ as a function of radial distance $r$, where $r = \sqrt{X^2+Y^2} = a (1- e \cos\epsilon)$, by splitting the disk into annular bins.

\bibliography{References}{}
\bibliographystyle{aasjournal}


\end{document}